# Large-Scale Model Selection with Misspecification *

Emre Demirkaya[1], Yang Feng[2], Pallavi Basu[3] and Jinchi Lv[1]

University of Southern California[1], Columbia University[2] and Tel Aviv University[3]

March 16, 2018

## Abstract

Model selection is crucial to high-dimensional learning and inference for contemporary big data applications in pinpointing the best set of covariates among a sequence of candidate interpretable models. Most existing work assumes implicitly that the models are correctly specified or have fixed dimensionality. Yet both features of model misspecification and high dimensionality are prevalent in practice. In this paper, we exploit the framework of model selection principles in misspecified models originated in Lv and Liu (2014) and investigate the asymptotic expansion of Bayesian principle of model selection in the setting of high-dimensional misspecified models. With a natural choice of prior probabilities that encourages interpretability and incorporates Kullback-Leibler divergence, we suggest the high-dimensional generalized Bayesian information criterion with prior probability (HGBIC$_p$) for large-scale model selection with misspecification. Our new information criterion characterizes the impacts of both model misspecification and high dimensionality on model selection. We further establish the consistency of covariance contrast matrix estimation and the model selection consistency of HGBIC$_p$ in ultra-high dimensions under some mild regularity conditions. The advantages of our new method are supported by numerical studies.

*Running title*: Large-Scale Model Selection with Misspecification

*Emre Demirkaya is Ph.D. Candidate, Department of Mathematics, University of Southern California, Los Angeles, CA 90089 (E-mail: *demirkay@usc.edu*). Yang Feng is Associate Professor, Department of Statistics, Columbia University, New York, NY 10027 (Email: yang.feng@columbia.edu). Pallavi Basu is Postdoctoral Scholar, Department of Statistics and Operations Research, Tel Aviv University, Tel Aviv, Israel 69978 (Email: plvibasu.work@gmail.com). Jinchi Lv is McAlister Associate Professor in Business Administration, Data Sciences and Operations Department, Marshall School of Business, University of Southern California, Los Angeles, CA 90089 (E-mail: *jinchilv@marshall.usc.edu*). Emre Demirkaya and Yang Feng contributed equally to this work. This work was supported by NSF CAREER Award DMS-1554804, a grant from the Simons Foundation, and Adobe Data Science Research Award.

# 1  Introduction

With rapid advances of modern technology, big data of unprecedented size, such as genetic and proteomic data, fMRI and functional data, and panel data in economics and finance, are frequently encountered in many contemporary applications. In these applications, the dimensionality $p$ can be comparable to or even much larger than the sample size $n$. A key assumption that often makes large-scale learning and inference feasible is the sparsity of signals, meaning that only a small fraction of covariates contribute to the response when $p$ is large compared to $n$. High-dimensional modeling with dimensionality reduction and feature selection plays an important role in these problems [15, 5, 16]. A sparse modeling procedure typically produces a sequence of candidate interpretable models, each involving a possibly different subset of covariates. An important question is how to compare different models in high dimensions when models are possibly misspecified.

The problem of model selection has been studied extensively by many researchers in the past several decades. Among others, well-known model selection criteria include the Akaike information criterion (AIC) [1, 2] and Bayesian information criterion (BIC) [32], where the former is based on the Kullback-Leibler (KL) divergence principle of model selection and the latter is originated from the Bayesian principle of model selection. A great deal of work has been devoted to understanding and extending these model selection criteria to different model settings; see, for example, [4, 20, 25, 24, 8, 9, 27, 30, 11, 23]. The connections between the AIC and cross-validation have been investigated in [34, 21, 31] for various contexts. In particular, [19] showed that classical information criteria such as AIC and BIC can no longer be consistent for model selection in ultra-high dimensions and proposed the generalized information criterion (GIC) for tuning parameter selection in high-dimensional penalized likelihood, for the scenario of correctly specified models. See also [3, 6, 7, 33, 18, 17] for some recent work on high-dimensional inference for feature selection.

Most existing work on model selection and feature selection usually makes an implicit assumption that the model under study is correctly specified or of fixed dimensions. Given the practical importance of model misspecification, [36] laid out a general theory of maximum likelihood estimation in misspecified models for the case of fixed dimensionality and independent and identically distributed (i.i.d.) observations. [10] also studied maximum likelihood estimation of a multi-dimensional log-concave density when the model is misspecified. Recently, [28] investigated the problem of model selection with model misspecification and originated asymptotic expansions of both KL divergence and Bayesian principles in misspecified generalized linear models, leading to the generalized AIC (GAIC) and generalized BIC (GBIC), for the case of fixed dimensionality. A specific form of prior probabilities motivated by the KL divergence principle led to the generalized BIC with prior probability (GBIC$_p$). Yet both features of model misspecification and high dimensionality are prevalent in contemporary big data applications. Thus an important question is how to characterize the impacts of both model misspecification and high dimensionality on model selection. We intend to provide some answer to this question in this paper.

Let us first gain some insights into the challenges of the aforementioned problem by considering a motivating example. Assume that the response $Y$ depends on the covariate vector $(X_1, \cdots, X_p)^T$ through the functional form

$$Y = f(X_1) + f(X_2 - X_3) + f(X_4 - X_5) + \varepsilon, \tag{1}$$

where $f(x) = x^3/(x^2 + 1)$ and the remaining setting is the same as in Section 4.1. Consider sample size $n = 200$ and vary dimensionality $p$ from 100 to 3200. Without any prior knowledge of the true model structure, we take the linear regression model

$$\mathbf{y} = \mathbf{Z}\boldsymbol{\beta} + \boldsymbol{\varepsilon} \tag{2}$$

as the working model and apply some information criteria to hopefully recover the oracle working model consisting of the first five covariates, where $\mathbf{y}$ is an $n$-dimensional response vector, $\mathbf{Z}$ is an $n \times p$ design matrix, $\boldsymbol{\beta} = (\beta_1, \cdots, \beta_p)^T$ is a $p$-dimensional regression coefficient vector, and $\boldsymbol{\varepsilon}$ is an $n$-dimensional error vector. When $p = 100$, the traditional AIC and BIC, which ignore model misspecification, tend to select a model with size larger than five. As expected, GBIC$_p$ in [28] works well by selecting the oracle working model over 90% of the time. However, when $p$ is increased to 3200, these methods fail to select such a model with significant probability and the prediction performance of the selected models deteriorates. This motivates us to study the problem of model selection in high-dimensional misspecified models. In contrast, our new method can recover the oracle working model with significant probability in this challenging scenario.

The main contributions of our paper are threefold. First, we provide the asymptotic expansion of Bayesian principle of model selection in high-dimensional misspecified generalized linear models which involves delicate and challenging technical analysis. Motivated by the asymptotic expansion and a natural choice of prior probabilities that encourages interpretability and incorporates Kullback-Leibler divergence, we suggest the high-dimensional generalized BIC with prior probability (HGBIC$_p$) for large-scale model selection with misspecification. Second, our work provides rigorous theoretical justification of the covariance contrast matrix estimator that incorporates the effect of model misspecification and is crucial for practical implementation. Such an estimator is shown to be consistent in the general setting of high-dimensional misspecified models. Third, we establish the model selection consistency of our new information criterion HGBIC$_p$ in ultra-high dimensions under some mild regularity conditions. In particular, our work provides important extensions to the studies in [28] and [19] to the cases of high dimensionality and model misspecification, respectively. The aforementioned contributions make our work distinct from other studies on model misspecification including [6, 23, 33].

The rest of the paper is organized as follows. Section 2 introduces the setup for model misspecification and presents the new information criterion HGBIC$_p$ based on Bayesian principle of model selection. We establish a systematic asymptotic theory for the new method in Section 3. Section 4 presents several numerical examples to demonstrate the advantages of our newly suggested method for large-scale model selection with misspecification. We provide some discussions of our results and possible extensions in Section 5. The proofs of main results are relegated to the Appendix. Additional technical details are provided in the Supplementary Material.

## 2 Large-scale model selection with misspecification

### 2.1 Model misspecification

The main focus of this paper is investigating ultra-high dimensional model selection with model misspecification in which the dimensionality $p$ can grow nonpolynomially with sample size $n$. We denote by $\mathfrak{M}$ an arbitrary subset with size $d$ of all $p$ available covariates and $\mathbf{X} = (\mathbf{x}_1, \cdots, \mathbf{x}_n)^T$ the corresponding $n \times d$ fixed design matrix given by the covariates in model $\mathfrak{M}$. Assume that conditional on the covariates in model $\mathfrak{M}$, the response vector $\mathbf{Y} = (Y_1, \cdots, Y_n)^T$ has independent components and each $Y_i$ follows distribution $G_{n,i}$ with density $g_{n,i}$, with all the distributions $G_{n,i}$ unknown to us in practice. Denote by $g_n = \prod_{i=1}^n g_{n,i}$ the product density and $G_n$ the corresponding *true* distribution of the response vector $\mathbf{Y}$.

Since the collection of true distributions $\{G_{n,i}\}_{1 \leq i \leq n}$ is unknown to practitioners, one often chooses a family of working models to fit the data. One class of popular working models is the family of generalized linear models (GLMs) [29] with a canonical link and natural parameter vector $\boldsymbol{\theta} = (\theta_1, \cdots, \theta_n)^T$ with $\theta_i = \mathbf{x}_i^T \boldsymbol{\beta}$, where $\mathbf{x}_i$ is a $d$-dimensional covariate vector and $\boldsymbol{\beta} = (\beta_1, \cdots, \beta_d)^T$ is a $d$-dimensional regression coefficient vector. Let $\tau > 0$ be the dispersion parameter. Then under the working model of GLM, the conditional density of response $y_i$ given the covariates in model $\mathfrak{M}$ is assumed to take the form

$$f_{n,i}(y_i) = \exp\{y_i \theta_i - b(\theta_i) + c(y_i, \tau)\}, \qquad (3)$$

where $b(\cdot)$ and $c(\cdot, \cdot)$ are some known functions with $b(\cdot)$ twice differentiable and $b''(\cdot)$ bounded away from 0 and $\infty$. $F_n$ denotes the corresponding distribution of the $n$-dimensional response vector $\mathbf{y} = (y_1, \cdots, y_n)^T$ with the product density $f_n = \prod_{i=1}^n f_{n,i}$ assuming the independence of components. Since the GLM is chosen by the user, the *working* distribution $F_n$ can be generally different from the true unknown distribution $G_n$.

For the GLM in (3) with natural parameter vector $\boldsymbol{\theta}$, let us define two vector-valued functions $\mathbf{b}(\boldsymbol{\theta}) = (b(\theta_1), \cdots, b(\theta_n))^T$ and $\boldsymbol{\mu}(\boldsymbol{\theta}) = (b'(\theta_1), \cdots, b'(\theta_n))^T$, and a matrix-valued function $\boldsymbol{\Sigma}(\boldsymbol{\theta}) = \text{diag}\{b''(\theta_1), \cdots, b''(\theta_n)\}$. The basic properties of GLM give the mean vector $E\mathbf{y} = \boldsymbol{\mu}(\boldsymbol{\theta})$ and the covariance matrix $\text{cov}(\mathbf{y}) = \boldsymbol{\Sigma}(\boldsymbol{\theta})$ with $\boldsymbol{\theta} = \mathbf{X}\boldsymbol{\beta}$. The product density of the response vector $\mathbf{y}$ can be written as

$$f_n(\mathbf{y};\boldsymbol{\beta},\tau) = \prod_{i=1}^{n} f_{n,i}(y_i) = \exp\left[\mathbf{y}^T\mathbf{X}\boldsymbol{\beta} - \mathbf{1}^T\mathbf{b}(\mathbf{X}\boldsymbol{\beta}) + \sum_{i=1}^{n} c(y_i,\tau)\right], \quad (4)$$

where $\mathbf{1}$ represents the $n$-dimensional vector with all components being one. Since GLM is only our working model, (4) results in the quasi-log-likelihood function [36]

$$\ell_n(\mathbf{y};\boldsymbol{\beta},\tau) = \log f_n(\mathbf{y};\boldsymbol{\beta},\tau) = \mathbf{y}^T\mathbf{X}\boldsymbol{\beta} - \mathbf{1}^T\mathbf{b}(\mathbf{X}\boldsymbol{\beta}) + \sum_{i=1}^{n} c(y_i,\tau). \quad (5)$$

Hereafter we treat the dispersion parameter $\tau$ as a known parameter and focus on our main parameter of interest $\boldsymbol{\beta}$. Whenever there is no confusion, we will slightly abuse the notation and drop the functional dependence on $\tau$.

The quasi-maximum likelihood estimator (QMLE) for the parameter vector $\boldsymbol{\beta}$ in our working model of GLM (3) is defined as

$$\widehat{\boldsymbol{\beta}}_n = \arg\max_{\boldsymbol{\beta}\in\mathbb{R}^d} \ell_n(\mathbf{y},\boldsymbol{\beta}), \quad (6)$$

which is the solution to the score equation

$$\boldsymbol{\Psi}_n(\boldsymbol{\beta}) = \partial\ell_n(\mathbf{y},\boldsymbol{\beta})/\partial\boldsymbol{\beta} = \mathbf{X}^T[\mathbf{y} - \boldsymbol{\mu}(\mathbf{X}\boldsymbol{\beta})] = \mathbf{0}. \quad (7)$$

For the linear regression model with $\boldsymbol{\mu}(\mathbf{X}\boldsymbol{\beta}) = \mathbf{X}\boldsymbol{\beta}$, such a score equation becomes the familiar normal equation $\mathbf{X}^T\mathbf{y} = \mathbf{X}^T\mathbf{X}\boldsymbol{\beta}$. Note that the KL divergence [26] of our working model $F_n$ from the true model $G_n$ is defined as $I(g_n; f_n(\cdot,\boldsymbol{\beta})) = E\log g_n(\mathbf{Y}) - E\ell_n(\mathbf{Y},\boldsymbol{\beta})$ with the response vector $\mathbf{Y}$ following the true distribution $G_n$. As in [28], we consider the best working model that is closest to the true model under the KL divergence. Such a model has parameter vector $\boldsymbol{\beta}_{n,0} = \arg\min_{\boldsymbol{\beta}\in\mathbb{R}^d} I(g_n; f_n(\cdot,\boldsymbol{\beta}))$, which solves the equation

$$\mathbf{X}^T\left[E\mathbf{Y} - \boldsymbol{\mu}(\mathbf{X}\boldsymbol{\beta})\right] = \mathbf{0}. \quad (8)$$

We see that equation (8) is simply the population version of the score equation given in (7).

Following [28], we introduce two matrices that play a key role in model selection with model misspecification. Note that under the true distribution $G_n$, we have $\text{cov}\left(\mathbf{X}^T\mathbf{Y}\right) = \mathbf{X}^T\text{cov}(\mathbf{Y})\mathbf{X}$. Computing the score equation at $\boldsymbol{\beta}_{n,0}$, we define matrix $\mathbf{B}_n$ as

$$\mathbf{B}_n = \text{cov}\left[\boldsymbol{\Psi}_n(\boldsymbol{\beta}_{n,0})\right] = \text{cov}\left(\mathbf{X}^T\mathbf{Y}\right) = \mathbf{X}^T\text{cov}(\mathbf{Y})\mathbf{X} \quad (9)$$

with $\text{cov}(\mathbf{Y}) = \text{diag}\{\text{var}(Y_1),\cdots,\text{var}(Y_n)\}$ by the independence assumption and under the true model. Note that under the working model $F_n$, it holds that $\text{cov}\left(\mathbf{X}^T\mathbf{Y}\right) = \mathbf{X}^T\boldsymbol{\Sigma}(\mathbf{X}\boldsymbol{\beta})\mathbf{X}$. We then define matrix $\mathbf{A}_n(\boldsymbol{\beta})$ as

$$\mathbf{A}_n(\boldsymbol{\beta}) = \frac{\partial^2 I(g_n; f_n(\cdot,\boldsymbol{\beta}))}{\partial\boldsymbol{\beta}^2} = -E\left\{\frac{\partial^2\ell_n(\mathbf{Y},\boldsymbol{\beta})}{\partial\boldsymbol{\beta}^2}\right\} = \mathbf{X}^T\boldsymbol{\Sigma}(\mathbf{X}\boldsymbol{\beta})\mathbf{X}, \quad (10)$$

and denote by $\mathbf{A}_n = \mathbf{A}_n(\boldsymbol{\beta}_{n,0})$. Hence we see that matrices $\mathbf{A}_n$ and $\mathbf{B}_n$ are the covariance matrices of $\mathbf{X}^T\mathbf{Y}$ under the best working model $F_n(\boldsymbol{\beta}_{n,0})$ and the true model $G_n$, respectively. To account for the effect of model misspecification, we define the covariance contrast matrix $\mathbf{H}_n = \mathbf{A}_n^{-1}\mathbf{B}_n$ as revealed in [28]. Observe that $\mathbf{A}_n$ and $\mathbf{B}_n$ coincide when the best working model and the true model are the same. In this case, $\mathbf{H}_n$ is an identity matrix of size $d$.

## 2.2 High-dimensional generalized BIC with prior probability

Given a set of competing models $\{\mathfrak{M}_m : m = 1, \cdots, M\}$, a popular model selection procedure using Bayesian principle of model selection is to first put nonzero prior probability $\alpha_{\mathfrak{M}_m}$ on each model $\mathfrak{M}_m$, and then choose a prior distribution $\mu_{\mathfrak{M}_m}$ for the parameter vector in the corresponding model. Assume that the density function of $\mu_{\mathfrak{M}_m}$ is bounded in $\mathbb{R}^{\mathfrak{M}_m} = \mathbb{R}^{d_m}$ with $d_m = |\mathfrak{M}_m|$ and locally bounded away from zero throughout the domain. The Bayesian principle of model selection is to choose the most probable model *a posteriori*; that is, choose the model $\mathfrak{M}_{m_0}$ such that

$$m_0 = \arg\max_{m \in \{1,\cdots,M\}} S(\mathbf{y}, \mathfrak{M}_m; F_n), \tag{11}$$

where the log-marginal-likelihood is

$$S(\mathbf{y}, \mathfrak{M}_m; F_n) = \log \int \alpha_{\mathfrak{M}_m} \exp\left[\ell_n(\mathbf{y}, \boldsymbol{\beta})\right] d\mu_{\mathfrak{M}_m}(\boldsymbol{\beta}) \tag{12}$$

with the log-likelihood $\ell_n(\mathbf{y}, \boldsymbol{\beta})$ as defined in (5) and the integral over $\mathbb{R}^{d_m}$.

The choice of prior probabilities $\alpha_{\mathfrak{M}_m}$ is important in high dimensions. [28] suggested the use of prior probability $\alpha_{\mathfrak{M}_m} \propto e^{-D_m}$ for each candidate model $\mathfrak{M}_m$, where the quantity $D_m$ is defined as

$$D_m = E\left[I(g_n; f_n(\cdot, \widehat{\boldsymbol{\beta}}_{n,m})) - I(g_n; f_n(\cdot, \boldsymbol{\beta}_{n,m,0}))\right] \tag{13}$$

with the subscript $m$ indicating a particular candidate model. The motivation is that the further the QMLE $\widehat{\boldsymbol{\beta}}_{n,m}$ is away from the best misspecified GLM $F_n(\cdot, \boldsymbol{\beta}_{n,m,0})$, the lower prior probability we assign to that model. In the high-dimensional setting when dimensionality $p$ can be much larger than sample size $n$, it is sensible to also take into account the complexity of the space of all possible sparse models with the same size as $\mathfrak{M}_m$. Such an observation motivates us to consider a new prior probability of the form

$$\alpha_{\mathfrak{M}_m} \propto p^{-d} e^{-D_m} \tag{14}$$

with $d = |\mathfrak{M}_m|$. The complexity factor $p^{-d}$ is motivated by the asymptotic expansion of $\left(\binom{p}{d} d!\right)^{-1}$. In fact, an application of Stirling's formula yields

$$\log\left(\binom{p}{d} d!\right)^{-1} \approx -d\log p = \log(p^{-d})$$

up to an additive term of order $o(d)$ when $d = o(p)$. The factor of $\binom{p}{d}^{-1}$ was also exploited in [8] who showed that using the term $\binom{p}{d}^{-\gamma}$ with some constant $0 < \gamma \leq 1$, the extended BIC can be model selection consistent for the scenario of correctly specified models with $p = O(n^\kappa)$ for some positive constant $\kappa$ satisfying $1 - (2\kappa)^{-1} < \gamma$. Moreover, we add the term $d!$ to reflect a stronger prior on model sparsity. See also [19] for the characterization of model selection in ultra-high dimensions with correctly specified models.

The asymptotic expansion of the Bayes factor for Bayesian principle of model selection in Theorem 1 to be presented in Section 3.2 motivates us to introduce the high-dimensional generalized BIC with prior probability (HGBIC$_p$) as follows for large-scale model selection with misspecification.

**Definition 1.** *We define* $\mathrm{HGBIC}_p = \mathrm{HGBIC}_p(\mathbf{y}, \mathfrak{M}; F_n)$ *of model* $\mathfrak{M}$ *as*

$$\mathrm{HGBIC}_p = -2\ell_n(\mathbf{y}, \widehat{\boldsymbol{\beta}}_n) + 2(\log p^*)|\mathfrak{M}| + \mathrm{tr}(\widehat{\mathbf{H}}_n) - \log |\widehat{\mathbf{H}}_n|, \tag{15}$$

*where* $\widehat{\mathbf{H}}_n$ *is a consistent estimator of* $\mathbf{H}_n$ *and* $p^* = pn^{1/2}$.

In correctly specified models, the term $\mathrm{tr}(\widehat{\mathbf{H}}_n) - \log |\widehat{\mathbf{H}}_n|$ in (15) is asymptotically close to $|\mathfrak{M}|$ when $\widehat{\mathbf{H}}_n$ is a consistent estimator of $\mathbf{H}_n = \mathbf{A}_n^{-1}\mathbf{B}_n = \mathbf{I}_d$. Thus compared to BIC with factor $\log n$, the HGBIC$_p$ contains a larger factor of order $\log p$ when dimensionality $p$ grows nonpolynomially with sample size $n$. This leads to a heavier penalty on model complexity similarly as in [19]. As shown in [28] for GBIC$_p$, the HGBIC$_p$ defined in (15) can also be viewed as a sum of three terms: the goodness of fit, model complexity, and model misspecification; see [28] for more details. Our new information criterion HGBIC$_p$ provides an important extension of the original GBIC$_p$ in [28], which was proposed for the scenario of model misspecification with fixed dimensionality, by explicitly taking into account the high dimensionality of the whole feature space.

## 3 Asymptotic properties of HGBIC$_p$

### 3.1 Technical conditions

We list the technical conditions required to prove the main results and the asymptotic properties of QMLE with diverging dimensionality. Denote by $\mathbf{Z}$ the full design matrix of size $n \times p$ whose $(i,j)$th entry is $x_{ij}$. For any subset $\mathfrak{M}$ of $\{1, \cdots, p\}$, $\mathbf{Z}_{\mathfrak{M}}$ denotes the submatrix of $\mathbf{Z}$ formed by columns whose indices are in $\mathfrak{M}$. When there is no confusion, we drop the subscript and use $\mathbf{X} = \mathbf{Z}_{\mathfrak{M}}$ for fixed $\mathfrak{M}$. For theoretical reasons, we restrict the parameter space to $\mathcal{B}_0$ which is a sufficiently large convex and compact set of $\mathbb{R}^p$. We consider parameters with bounded support. Namely, we define $\mathcal{B}(\mathfrak{M}) = \{\boldsymbol{\beta} \in \mathcal{B}_0 : \mathrm{supp}(\boldsymbol{\beta}) = \mathfrak{M}\}$ and $\mathcal{B} = \cup_{|\mathfrak{M}| \leq K} \mathcal{B}(\mathfrak{M})$ where the maximum support size $K$ is taken to be $o(n)$. Moreover, we assume that $c_0 \leq b''(\mathbf{Z}\boldsymbol{\beta}) \leq c_0^{-1}$ for any $\boldsymbol{\beta} \in \mathcal{B}$ where $c_0$ is some positive constant.

We use the following notation. For matrices, $\|\cdot\|_2$, $\|\cdot\|_\infty$, and $\|\cdot\|_F$ denote the matrix operator norm, entrywise maximum norm, and matrix Frobenius norm, respectively. For vectors, $\|\cdot\|_2$ and $\|\cdot\|_\infty$ denote the vector $L_2$-norm and maximum norm, and $(\mathbf{v})_i$ represents the $i$th component of vector $\mathbf{v}$. Denote by $\lambda_{\min}(\cdot)$ and $\lambda_{\max}(\cdot)$ the smallest and largest eigenvalues of a given matrix, respectively.

**Condition 1.** *There exists some positive constant $c_1$ such that for each $1 \leq i \leq n$,*

$$P(|W_i| > t) \leq c_1 \exp(-c_1^{-1} t) \tag{16}$$

*for any $t > 0$, where $\mathbf{W} = (W_1, \cdots, W_n)^T = \mathbf{Y} - E\mathbf{Y}$. The variances of $Y_i$ are bounded below uniformly in $i$ and $n$.*

**Condition 2.** *Let $u_1$ and $u_2$ be some positive constants and $m_n = O(n^{u_1})$ a diverging sequence. We have the following bounds*

*(i) $\max\{\|E\mathbf{Y}\|_\infty, \sup_{\boldsymbol{\beta} \in \mathcal{B}} \|\boldsymbol{\mu}(\mathbf{Z}\boldsymbol{\beta})\|_\infty\} \leq m_n$;*

*(ii) $\sum_{i=1}^n \left[\frac{[EY_i - (\boldsymbol{\mu}(\mathbf{X}\boldsymbol{\beta}_{n,0}))_i]^2}{\mathrm{var}(Y_i)}\right]^2 = O(n^{u_2})$.*

*For simplicity, we also assume that $m_n$ diverges faster than $\log n$.*

**Condition 3.** *Let $K = o(n)$ be a positive integer. There exist positive constants $c_2$ and $u_3$ such that*

*(i) For any $\mathfrak{M} \subset \{1, \cdots, p\}$ such that $|\mathfrak{M}| \leq K$,*

$$c_2 \leq \lambda_{\min}(n^{-1} \mathbf{Z}_\mathfrak{M}^T \mathbf{Z}_\mathfrak{M}) \leq \lambda_{\max}(n^{-1} \mathbf{Z}_\mathfrak{M}^T \mathbf{Z}_\mathfrak{M}) \leq c_2^{-1};$$

*(ii) $\|\mathbf{Z}\|_\infty = O(n^{u_3})$;*

*(iii) For simplicity, we assume that columns of $\mathbf{Z}$ are normalized: $\sum_{i=1}^n x_{ij}^2 = n$ for all $j = 1, \cdots, p$.*

Condition 1 is a standard tail condition on the response variable $Y$. This condition ensures that the sub-exponential norm of the response is bounded. Conditions 2 and 3 have their counterparts in [19]. However, Condition 2 is modified to deal with model misspecification. More specifically, the means of the true distribution and fitted model as well as their relations are assumed in Condition 2. The first part simultaneously controls the tail behavior of the response and fitted model. The second part ensures that the mean of the fitted distribution does not deviate from the true mean too significantly. Condition 3 is on the design matrix $\mathbf{X}$. The first part is important for the consistency of QMLE $\widehat{\boldsymbol{\beta}}_n$ and uniqueness of the population parameter. Conditions 2 and 3 also provide bounds for the eigenvalues of $\mathbf{A}_n(\boldsymbol{\beta})$ and $\mathbf{B}_n$. See [19] for further discussions on these assumptions.

For the following conditions, we define a neighborhood around $\boldsymbol{\beta}_{n,0}$. Let $\delta_n = m_n\sqrt{\log p} = O(n^{u_1}\sqrt{\log p})$. We define the neighborhood $N_n(\delta_n) = \{\boldsymbol{\beta} \in \mathbb{R}^d : \|(n^{-1}\mathbf{B}_n)^{1/2}(\boldsymbol{\beta} - \boldsymbol{\beta}_{n,0})\|_2 \leq (n/d)^{-1/2}\delta_n\}$. We assume that $(n/d)^{-1/2}\delta_n$ converges to zero so that $N_n(\delta_n)$ is an asymptotically shrinking neighborhood of $\boldsymbol{\beta}_{n,0}$.

**Condition 4.** *Assume that the prior density relative to the Lebesgue measure $\mu_0$ on $\mathbb{R}^d$ $\pi(h(\boldsymbol{\beta})) = \frac{d\mu_{\mathfrak{M}}}{d\mu_0}(h(\boldsymbol{\beta}))$ satisfies*

$$\inf_{\boldsymbol{\beta} \in N_n(2\delta_n)} \pi(h(\boldsymbol{\beta})) \geq c_3 \text{ and } \sup_{\boldsymbol{\beta} \in \mathbb{R}^d} \pi(h(\boldsymbol{\beta})) \leq c_3^{-1}, \tag{17}$$

*where $c_3$ is a positive constant, and $h(\boldsymbol{\beta}) = (n^{-1}\mathbf{B}_n)^{1/2}\boldsymbol{\beta}$.*

**Condition 5.** *Let $\mathbf{V}_n(\boldsymbol{\beta}) = \mathbf{B}_n^{-1/2}\mathbf{A}_n(\boldsymbol{\beta})\mathbf{B}_n^{-1/2}$, $\mathbf{V}_n = \mathbf{V}_n(\boldsymbol{\beta}_{n,0}) = \mathbf{B}_n^{-1/2}\mathbf{A}_n\mathbf{B}_n^{-1/2}$, and $\widetilde{\mathbf{V}}_n(\boldsymbol{\beta}_1, \cdots, \boldsymbol{\beta}_d) = \mathbf{B}_n^{-1/2}\widetilde{\mathbf{A}}_n(\boldsymbol{\beta}_1, \cdots, \boldsymbol{\beta}_d)\mathbf{B}_n^{-1/2}$, where $\widetilde{\mathbf{A}}_n(\boldsymbol{\beta}_1, \cdots, \boldsymbol{\beta}_d)$ is the matrix whose $j$th row is the corresponding row of $\mathbf{A}_n(\boldsymbol{\beta}_j)$ for each $j = 1, \cdots, d$. There exists some sequence $\rho_n(\delta_n)$ such that $\rho_n(\delta_n)\delta_n^2 d$ converges to zero,*

*(i)* $\max_{\boldsymbol{\beta} \in N_n(2\delta_n)} \max\{|\lambda_{\min}(\mathbf{V}_n(\boldsymbol{\beta}) - \mathbf{V}_n)|, |\lambda_{\max}(\mathbf{V}_n(\boldsymbol{\beta}) - \mathbf{V}_n)|\} \leq \rho_n(\delta_n);$

*(ii)* $\max_{\boldsymbol{\beta}_1, \cdots, \boldsymbol{\beta}_d \in N_n(\delta_n)} \|\widetilde{\mathbf{V}}_n(\boldsymbol{\beta}_1, \cdots, \boldsymbol{\beta}_d) - \mathbf{V}_n\|_2 \leq \rho_n(\delta_n).$

Similar versions of Conditions 4 and 5 were imposed in [28]. Under Condition 4, the prior density is bounded above globally and bounded below in a neighborhood of $\boldsymbol{\beta}_{n,0}$. This condition is used in Theorem 1 for the asymptotic expansion of the Bayes factor. Condition 5 is on the continuity of the matrix-valued function $\mathbf{V}_n$ and $\widetilde{\mathbf{V}}_n$ in a shrinking neighborhood $N_n(2\delta_n)$ of $\boldsymbol{\beta}_{n,0}$. The first and second parts control the expansions of expected log-likelihood and score functions, respectively. Condition 5 ensures that the remainders are negligible in approximating the log-marginal-likelihood $S(\mathbf{y}, \mathfrak{M}_m; F_n)$. See [28] for more discussions.

### 3.2 Asymptotic expansion of Bayesian principle of model selection

We now provide the asymptotic expansion of Bayesian principle of model selection with the prior introduced in Section 2.2. As mentioned earlier, the Bayesian principle chooses the model that maximizes the log-marginal-likelihood given in (12). To ease the presentation, for any $\boldsymbol{\beta} \in \mathbb{R}^d$, we define a quantity

$$\ell_n^*(\mathbf{y}, \boldsymbol{\beta}) = \ell_n(\mathbf{y}, \boldsymbol{\beta}) - \ell_n(\mathbf{y}, \widehat{\boldsymbol{\beta}}_n), \tag{18}$$

which is the deviation of the quasi-log-likelihood from its maximum. Then from (12) and (18), we have

$$S(\mathbf{y}, \mathfrak{M}_m; F_n) = \ell_n(\mathbf{y}, \widehat{\boldsymbol{\beta}}_n) + \log E_{\mu_{\mathfrak{M}_m}}[U_n(\boldsymbol{\beta})^n] + \log \alpha_{\mathfrak{M}_m}, \tag{19}$$

where $U_n(\boldsymbol{\beta}) = \exp[n^{-1}\ell_n^*(\mathbf{y}, \boldsymbol{\beta})]$. With the choice of the prior probability in (14), it is clear that

$$\log \alpha_{\mathfrak{M}_m} = -D_m - d \log p. \tag{20}$$

Aided by (19) and (20), some delicate technical analysis unveils the following expansion of the log-marginal-likelihood.

**Theorem 1.** *Let $\alpha_{\mathfrak{M}_m} = Cp^{-d}e^{-D_m}$ with $C > 0$ some normalization constant and assume that Conditions 1, 2(i), 3(i), 3(iii), 4, and 5 hold. If $(n/d)^{-1/2}\delta_n = o(1)$, then we have with probability tending to one,*

$$S(\mathbf{y}, \mathfrak{M}; F_n) = \ell_n(\mathbf{y}, \widehat{\boldsymbol{\beta}}_n) - (\log p^*)|\mathfrak{M}| - \frac{1}{2}\mathrm{tr}(\mathbf{H}_n) + \frac{1}{2}\log|\mathbf{H}_n|$$
$$+ \log(Cc_n) + o(\mu_n), \qquad (21)$$

*where $\mathbf{H}_n = \mathbf{A}_n^{-1}\mathbf{B}_n$, $p^* = pn^{1/2}$, $\mu_n = \mathrm{tr}(\mathbf{A}_n^{-1}\mathbf{B}_n) \vee 1$, and $c_n \in [c_3, c_3^{-1}]$.*

Theorem 1 lays the foundation for investigating high-dimensional model selection with model misspecification. Based on the asymptotic expansion in (21), our new information criterion HGBIC$_p$ in (15) is defined by replacing the covariance contrast matrix $\mathbf{H}_n$ with a consistent estimator $\widehat{\mathbf{H}}_n$. The HGBIC$_p$ naturally characterizes the impacts of both model misspecification and high dimensionality on model selection. A natural question is how to ensure a consistent estimator for $\mathbf{H}_n$. We address such a question in the next section.

### 3.3 Consistency of covariance contrast matrix estimation

For practical implementation of HGBIC$_p$, it is of vital importance to provide a consistent estimator for the covariance contrast matrix $\mathbf{H}_n$. To this end, we consider the plug-in estimator $\widehat{\mathbf{H}}_n = \widehat{\mathbf{A}}_n^{-1}\widehat{\mathbf{B}}_n$ with $\widehat{\mathbf{A}}_n$ and $\widehat{\mathbf{B}}_n$ defined as follows. Since the QMLE $\widehat{\boldsymbol{\beta}}_n$ provides a consistent estimator of $\boldsymbol{\beta}_{n,0}$ in the best misspecified GLM $F_n(\cdot, \boldsymbol{\beta}_{n,0})$, a natural estimate of matrix $\mathbf{A}_n$ is given by

$$\widehat{\mathbf{A}}_n = \mathbf{A}_n(\widehat{\boldsymbol{\beta}}_n) = \mathbf{X}^T\boldsymbol{\Sigma}(\mathbf{X}\widehat{\boldsymbol{\beta}}_n)\mathbf{X}. \qquad (22)$$

When the model is correctly specified, the following simple estimator

$$\widehat{\mathbf{B}}_n = \mathbf{X}^T\mathrm{diag}\left\{\left[\mathbf{y} - \boldsymbol{\mu}(\mathbf{X}\widehat{\boldsymbol{\beta}}_n)\right] \circ \left[\mathbf{y} - \boldsymbol{\mu}(\mathbf{X}\widehat{\boldsymbol{\beta}}_n)\right]\right\}\mathbf{X} \qquad (23)$$

with $\circ$ denoting the componentwise product gives an asymptotically unbiased estimator of matrix $\mathbf{B}_n$.

**Theorem 2.** *Assume that Conditions 1–3 hold, $n^{-1}\mathbf{A}_n(\boldsymbol{\beta})$ is Lipschitz in operator norm in the neighborhood $N_n(\delta_n)$, $d = O(n^{\kappa_1})$, and $\log p = O(n^{\kappa_2})$ with constants satisfying $0 < \kappa_1 < 1/4$, $0 < u_3 < 1/4 - \kappa_1$, $0 < u_2 < 1 - 4\kappa_1 - 4u_3$, $0 < u_1 < 1/2 - 2\kappa_1 - u_3$, and $0 < \kappa_2 < 1 - 4\kappa_1 - 2u_1 - 2u_3$. Then the plug-in estimator $\widehat{\mathbf{H}}_n = \widehat{\mathbf{A}}_n^{-1}\widehat{\mathbf{B}}_n$ satisfies that $\mathrm{tr}(\widehat{\mathbf{H}}_n) = \mathrm{tr}(\mathbf{H}_n) + o_P(1)$ and $\log|\widehat{\mathbf{H}}_n| = \log|\mathbf{H}_n| + o_P(1)$.*

Theorem 2 improves the result in [28] in two important aspects. First, the consistency of the covariance contrast matrix estimator was justified in [28] only for the scenario of correctly specified models. Our new result shows that the simple plug-in estimator $\widehat{\mathbf{H}}_n$ still enjoys consistency in the general setting of model misspecification. Second, the result in Theorem 2 holds for the case of high dimensionality. These theoretical guarantees are crucial to the practical implementation of the new information criterion HGBIC$_p$. Our numerical studies in Section 4 reveal that such an estimate works well in a variety of model misspecification settings.

### 3.4 Model selection consistency of HGBIC$_p$

We further investigate the model selection consistency property of information criterion HGBIC$_p$. Assume that there are $M = o(n^\delta)$ sparse candidate models $\mathfrak{M}_1, \cdots \mathfrak{M}_M$, where $\delta$ is some sufficiently large positive constant. For each candidate model $\mathfrak{M}_m$, we have the HGBIC$_p$ criterion as defined in (15)

$$\text{HGBIC}_p(\mathfrak{M}_m) = -2\ell_n(\mathbf{y}, \widehat{\boldsymbol{\beta}}_{n,m}) + 2(\log p^*)|\mathfrak{M}_m| + \text{tr}(\widehat{\mathbf{H}}_{n,m}) - \log |\widehat{\mathbf{H}}_{n,m}|, \qquad (24)$$

where $\widehat{\mathbf{H}}_{n,m}$ is a consistent estimator of $\mathbf{H}_{n,m}$ and $p^* = pn^{1/2}$. Assume that there exists an oracle working model in the sequence $\{\mathfrak{M}_m : m = 1, \cdots, M\}$ that has support identical to the set of all important features in the true model. Without loss of generality, suppose that $\mathfrak{M}_1$ is such oracle working model.

**Theorem 3.** *Assume that all the conditions of Theorems 1–2 hold and the population version of HGBIC$_p$ criterion in (24) is minimized at $\mathfrak{M}_1$ such that for some $\Delta > 0$,*

$$\min_{m>1} \left\{ \text{HGBIC}_p^*(\mathfrak{M}_m) - \text{HGBIC}_p^*(\mathfrak{M}_1) \right\} > \Delta \qquad (25)$$

*with* $\text{HGBIC}_p^*(\mathfrak{M}_m) = -2\ell_n(\mathbf{y}, \boldsymbol{\beta}_{n,m,0}) + 2(\log p^*)|\mathfrak{M}_m| + \text{tr}(\mathbf{H}_{n,m}) - \log |\mathbf{H}_{n,m}|$. *Then it holds that*

$$\min_{m>1} \left\{ \text{HGBIC}_p(\mathfrak{M}_m) - \text{HGBIC}_p(\mathfrak{M}_1) \right\} > \Delta/2 \qquad (26)$$

*with asymptotic probability one.*

Theorem 3 formally establishes the model selection consistency property of the new information criterion HGBIC$_p$ for large-scale model selection with misspecification in that the oracle working model can be selected among a large sequence of candidate sparse models with significant probability. Such a desired property is an important consequence of results in Theorems 1 and 2.

## 4 Numerical studies

We now investigate the finite-sample performance of the information criterion HGBIC$_p$ in comparison to the information criteria AIC, BIC, GAIC, GBIC, and GBIC$_p$ in high-dimensional misspecified models via simulation examples.

## 4.1 Multiple index model

The first model we consider is the following multiple index model

$$Y = f(\beta_1 X_1) + f(\beta_2 X_2 + \beta_3 X_3) + f(\beta_4 X_4 + \beta_5 X_5) + \varepsilon, \qquad (27)$$

where the response depends on the covariates $X_j$'s only through the first five ones in a nonlinear fashion and $f(x) = x^3/(x^2 + 1)$. Here the rows of the $n \times p$ design matrix $\mathbf{Z}$ are sampled as i.i.d. copies from $N(\mathbf{0}, I_p)$, and the $n$-dimensional error vector $\boldsymbol{\varepsilon} \sim N(\mathbf{0}, \sigma^2 I_n)$. We set the true parameter vector $\boldsymbol{\beta}_0 = (1, -1, 1, 1, -1, 0, \cdots, 0)^T$ and $\sigma = 0.8$. We vary the dimensionality $p$ from 100 to 3200 while keeping the sample size $n$ fixed at 200. We would like to investigate the behavior of different information criteria when the dimensionality increases. Although the data was generated from model (27), we fit the linear regression model (2). This is a typical example of model misspecification. Note that since the first five variables are independent of the other variables, the oracle working model is $M_0 = \text{supp}(\boldsymbol{\beta}_0) = \{1, \cdots, 5\}$. Due to the high dimensionality, it is computationally prohibitive to implement the best subset selection. Thus we first applied Lasso followed by least-squares refitting to build a sequence of sparse models and then selected the final model using a model selection criterion. In practice, one can apply any preferred variable selection procedure to obtain a sequence of candidate interpretable models.

We report the consistent selection probability (the proportion of simulations where selected model $\widehat{M} = M_0$), the sure screening probability [14, 13] (the proportion of simulations where selected mode $\widehat{M} \supset M_0$), and the prediction error $E(Y - \mathbf{z}^T \widehat{\boldsymbol{\beta}})^2$ with $\widehat{\boldsymbol{\beta}}$ an estimate and $(\mathbf{z}, Y)$ an independent observation for $\mathbf{z} = (X_1, \cdots, X_p)^T$. To evaluate the prediction performance of different criteria, we calculated the average prediction error on an independent test sample of size 10,000. The results for prediction error and model selection performance are summarized in Table 1. In addition, we calculate the average number of false positives for each method in Table 2.

From Table 1, we observe that as the dimensionality $p$ increases, the consistent selection probability tends to decrease for all criteria except the newly suggested HGBIC$_p$, which maintains a perfect 100% consistent selection probability throughout all dimensions considered. Generally speaking, GAIC improved over AIC, and GBIC, GBIC$_p$ performed better than BIC in terms of both prediction and variable selection. In particular, the model selected by our new information criterion HGBIC$_p$ delivered the best performance with the smallest prediction error and highest consistent selection probability across all settings.

An interesting observation is the comparison between GBIC$_p$ and HGBIC$_p$ in terms of model selection consistency property. While GBIC$_p$ is comparable to HGBIC$_p$ when the dimensionality is not large (e.g., $p = 100$ and 200), the difference between these two methods increases as the dimensionality increases. In the case when $p = 3200$, HGBIC$_p$ has 100% of success for consistent selection, while GBIC$_p$ has success rate of only 45%. This confirms the necessity of including the $\log p^*$ factor with $p^* = pn^{1/2}$ in the model selection criterion to take into account the high dimensionality, which is in line with the results in [19] for the case of correctly specified models.

Table 1: Average results over 100 repetitions for Example 4.1 with all entries multiplied by 100.

| $p$ | AIC | BIC | GAIC | GBIC | GBIC$_p$ | HGBIC$_p$ | Oracle |
|---|---|---|---|---|---|---|---|
| Consistent selection probability with sure screening probability in parentheses | | | | | | | |
| 100 | 1(100) | 71(100) | 5(100) | 72(100) | 92(100) | 100(100) | 100(100) |
| 200 | 0(100) | 43(100) | 4(100) | 44(100) | 79(100) | 100(100) | 100(100) |
| 400 | 0(100) | 27(100) | 1(100) | 31(100) | 67(100) | 100(100) | 100(100) |
| 800 | 0(100) | 16(100) | 1(100) | 21(100) | 57(100) | 100(100) | 100(100) |
| 1600 | 0(100) | 13(100) | 2(100) | 18(100) | 49(100) | 100(100) | 100(100) |
| 3200 | 0(100) | 5(100) | 3(100) | 8(100) | 45(100) | 100(100) | 100(100) |
| Mean prediction error with standard error in parentheses | | | | | | | |
| 100 | 97(1) | 84(1) | 89(1) | 84(1) | 82(1) | 82(1) | 82(1) |
| 200 | 103(1) | 84(1) | 89(1) | 84(1) | 81(1) | 80(1) | 80(1) |
| 400 | 112(2) | 88(1) | 94(1) | 87(1) | 84(1) | 82(1) | 82(1) |
| 800 | 120(1) | 93(1) | 97(1) | 92(1) | 86(1) | 83(1) | 83(1) |
| 1600 | 121(1) | 94(1) | 96(1) | 93(1) | 87(1) | 82(1) | 82(1) |
| 3200 | 117(1) | 93(1) | 93(1) | 91(1) | 84(1) | 80(1) | 80(1) |

We further study a family of model selection criteria induced by the HGBIC$_p$ and characterized as follows

$$\text{HGBIC}_{p,\zeta}(\mathfrak{M}_m) = -2\ell_n(\mathbf{y}, \widehat{\boldsymbol{\beta}}_{n,m}) + \zeta \left[ 2(\log p^*)|\mathfrak{M}_m| + \text{tr}(\widehat{\mathbf{H}}_{n,m}) - \log|\widehat{\mathbf{H}}_{n,m}| \right], \qquad (28)$$

where $\zeta$ is a positive factor controlling the penalty level on both model misspecification and high dimensionality. Note that HGBIC$_{p,\zeta}$ with $\zeta = 1$ reduces to our original HGBIC$_p$. Here we examine the impact of the factor $\zeta$ on the false discovery proportion (FDP) and the true positive rate (TPR) for the selected model $\widehat{M}$ compared to the oracle working model $M_0$. In Figure 1, we observe that as $\zeta$ increases, the average FDP drops sharply as it gets close to 1. In addition, we have the desired model selection consistency property (with FDP close to 0 and TPR close to 1) when $\zeta \in [1, 2]$. This figure demonstrates the robustness of the introduced HGBIC$_{p,\zeta}$ criteria.

Table 2: Average false positives over 100 repetitions for Example 4.1.

|  | AIC | BIC | GAIC | GBIC | GBIC$_p$ | HGBIC$_p$ |
|---|---|---|---|---|---|---|
| 100 | 8.55 | 0.51 | 3.49 | 0.48 | 0.08 | 0.00 |
| 200 | 13.05 | 1.07 | 3.70 | 0.98 | 0.28 | 0.00 |
| 400 | 17.68 | 1.65 | 4.66 | 1.49 | 0.40 | 0.00 |
| 800 | 21.28 | 2.61 | 4.57 | 2.17 | 0.70 | 0.00 |
| 1600 | 22.20 | 3.01 | 4.40 | 2.68 | 0.96 | 0.00 |
| 3200 | 22.48 | 3.92 | 4.07 | 3.20 | 0.86 | 0.00 |

Figure 1: The average false discovery proportion (left panel) and the true positive rate (right panel) as the factor $\zeta$ varies for Example 4.1.

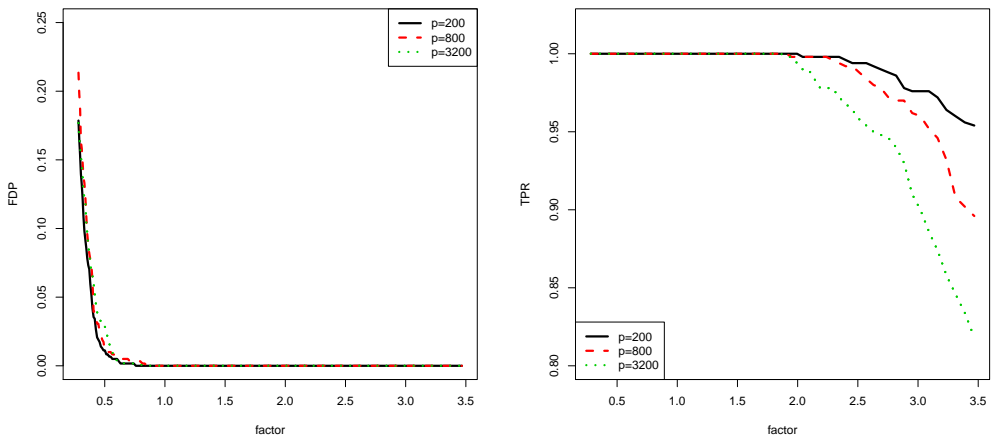

### 4.2 Logistic regression with interaction

Our second simulation example is the high-dimensional logistic regression with interaction. We simulated 100 data sets from the logistic regression model with interaction and an $n$-dimensional parameter vector

$$\boldsymbol{\theta} = \mathbf{Z}\boldsymbol{\beta} + 2\mathbf{x}_{p+1} + 2\mathbf{x}_{p+2}, \tag{29}$$

where $\mathbf{Z} = (\mathbf{x}_1, \cdots, \mathbf{x}_p)$ is an $n \times p$ design matrix, $\mathbf{x}_{p+1} = \mathbf{x}_1 \circ \mathbf{x}_2$ and $\mathbf{x}_{p+2} = \mathbf{x}_3 \circ \mathbf{x}_4$ are two interaction terms, and the rest of the setting is the same as in the simulation example in Section 4.1. For each data set, the $n$-dimensional response vector $\mathbf{y}$ was sampled from the Bernoulli distribution with success probability vector $[e^{\theta_1}/(1+e^{\theta_1}), \cdots, e^{\theta_n}/(1+e^{\theta_n})]^T$ with $\boldsymbol{\theta} = (\theta_1, \cdots, \theta_n)^T$ given in (29). As in Section 4.1, we consider the case where all covariates are independent of each other. We chose $\boldsymbol{\beta}_0 = (2.5, -1.9, 2.8, -2.2, 3, 0, \cdots, 0)^T$ and set sample size $n = 300$. Although the data was generated from the logistic regression model with parameter vector (29), we fit the logistic regression model without the two interaction terms. This provides another example of misspecified models. As argued in Section 4.1, the oracle working model is $\text{supp}(\boldsymbol{\beta}_0) = \{1, \cdots, 5\}$ which corresponds to the logistic regression model with the first five covariates. To build a sequence of sparse models, we applied Lasso followed by maximum-likelihood refitting based on the support of the estimated model.

Table 3: Average results over 100 repetitions for Example 4.2 with all entries multiplied by 100.

| $p$ | AIC | BIC | GAIC | GBIC | $\text{GBIC}_p$ | $\text{HGBIC}_p$ | Oracle |
|---|---|---|---|---|---|---|---|
| Consistent selection probability with sure screening probability probability in parentheses | | | | | | | |
| 100 | 0(100) | 30(100) | 0(100) | 32(100) | 55(100) | 99(99) | 100(100) |
| 200 | 0(100) | 27(100) | 0(100) | 29(100) | 41(100) | 95(97) | 100(100) |
| 400 | 0(100) | 12(100) | 0(100) | 16(100) | 35(100) | 95(95) | 100(100) |
| 800 | 0(100) | 2(99) | 0(100) | 4(99) | 12(99) | 94(95) | 100(100) |
| 1600 | 0(100) | 0(100) | 0(100) | 1(100) | 5(100) | 84(85) | 100(100) |
| 3200 | 0(100) | 0(100) | 0(100) | 1(100) | 1(100) | 79(84) | 100(100) |
| Mean classification error (in percentage) with standard error in parentheses | | | | | | | |
| 100 | 25.3(0.3) | 16.1(0.2) | 27.4(0.3) | 16.1(0.2) | 15.7(0.2) | 15.2(0.2) | 15.2(0.2) |
| 200 | 24.9(0.3) | 17.2(0.3) | 28.4(0.3) | 16.9(0.3) | 16.5(0.2) | 15.5(0.2) | 15.4(0.2) |
| 400 | 25.0(0.3) | 19.7(0.4) | 28.7(0.3) | 17.8(0.3) | 16.8(0.3) | 15.3(0.2) | 15.2(0.2) |
| 800 | 24.7(0.3) | 21.9(0.4) | 28.0(0.3) | 18.8(0.3) | 17.7(0.3) | 15.7(0.2) | 15.5(0.2) |
| 1600 | 26.0(0.4) | 24.3(0.4) | 28.4(0.3) | 20.2(0.3) | 18.7(0.3) | 15.9(0.3) | 15.4(0.2) |
| 3200 | 25.7(0.3) | 24.4(0.4) | 28.2(0.3) | 20.7(0.3) | 19.5(0.3) | 15.9(0.2) | 15.3(0.2) |

Since the goal in logistic regression is usually classification, we replace the prediction error with the classification error rate. Tables 3 and 4 show similar conclusions to those in Section 4.1. Again $\text{HGBIC}_p$ outperformed all other model selection criteria with greater advantage as the dimensionality increases (e.g., $p \geq 800$). As in Example 4.1, we also present the trend of FDP and TPR as $\zeta$ varies in Figure 2. From the figure, we observe that it is a more difficult setting than the multiple index model to reach model selection consistency. The proposed $\text{HGBIC}_p$ criterion with the choice of $\zeta = 1$ appears to strike a good balance between FDP and TPR.

Table 4: Average false positives over 100 repetitions for Example 4.2.

|      | AIC   | BIC   | GAIC   | GBIC | $\text{GBIC}_p$ | $\text{HGBIC}_p$ |
|------|-------|-------|--------|------|------|-------|
| 100  | 55.71 | 1.57  | 86.50  | 1.48 | 0.78 | 0.00  |
| 200  | 40.83 | 3.24  | 119.82 | 2.14 | 1.33 | 0.02  |
| 400  | 35.25 | 11.74 | 130.33 | 4.27 | 2.24 | 0.00  |
| 800  | 31.78 | 18.22 | 140.20 | 6.00 | 3.45 | 0.01  |
| 1600 | 30.25 | 22.65 | 142.81 | 8.02 | 4.80 | 0.01  |
| 3200 | 28.41 | 22.31 | 142.75 | 8.61 | 6.15 | 0.05  |

Figure 2: The average false discovery proportion (left panel) and the true positive rate (right panel) as the factor $\zeta$ varies for Example 4.2.

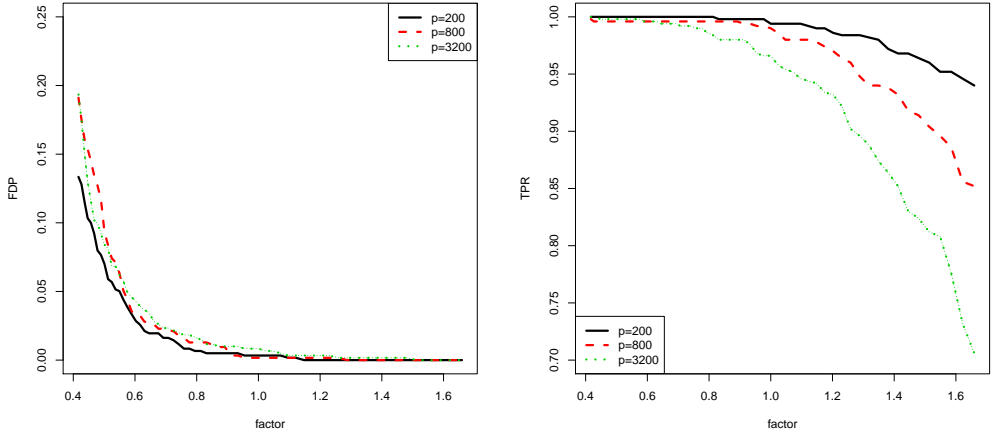

## 5 Discussions

Despite the rich literature on model selection, the general case of model misspecification in high dimensions is less well studied. Our work has investigated the problem of model selection in high-dimensional misspecified models and characterized the impacts of both model misspecification and high dimensionality on model selection, providing an important extension of the work in [28] and [19]. The newly suggested information criterion $\text{HGBIC}_p$ has been shown to perform well in high-dimensional settings. Moreover, we have established the consistency of the covariance contrast matrix estimator that captures the effect of model misspecification in the general setting, and the model selection consistency of $\text{HGBIC}_p$ in ultra-high dimensions.

The $\log p^*$ term in $\text{HGBIC}_p$ with $p^* = pn^{1/2}$ is adaptive to high dimensions. In the setting of correctly specified models, [19] showed that a similar term is necessary for the model selection consistency of information criteria when the dimensionality $p$ grows fast with the sample size $n$. It would be interesting to study optimality property of the information criteria HGBIC$_p$ and the HGBIC$_{p,\zeta}$ defined in (28) under model misspecification, and investigate these model selection principles in more general high-dimensional misspecified models such as the additive models and survival models. It would also be interesting to combine the strengths of the newly suggested HGBIC$_p$ and the recently introduced knockoffs inference framework [3, 7, 17] for more stable and enhanced large-scale model selection with misspecification. These problems are beyond the scope of the current paper and are interesting topics for future research.

## A  Proofs of main results

We provide the proofs of Theorems 1–3 in this appendix. Additional technical details are provided in the Supplementary Material.

### A.1  Proof of Theorem 1

We consider the decomposition of $S(\mathbf{y}, \mathfrak{M}_m; F_n)$ in (19) and deal with terms $\log E_{\mu_{\mathfrak{M}_m}}[U_n(\boldsymbol{\beta})^n]$ and $\log \alpha_{\mathfrak{M}_m}$ separately by invoking Taylor's expansion. In fact, $\log E_{\mu_{\mathfrak{M}_m}}[U_n(\boldsymbol{\beta})^n]$ is based on $\ell_n^*(\mathbf{y}, \boldsymbol{\beta})$, the deviation of the quasi-log-likelihood from its maximum, while $\log \alpha_{\mathfrak{M}_m}$ is the log-prior probability which depends on $D_m = E[I(g_n; f_n(\cdot, \widehat{\boldsymbol{\beta}}_{n,m})) - I(g_n; f_n(\cdot, \boldsymbol{\beta}_{n,m,0}))]$, expected difference in the KL divergences. In light of consistency of the estimator $\widehat{\boldsymbol{\beta}}_n$ as shown in Lemma 1, we focus only on the neighborhood of $\boldsymbol{\beta}_{n,0}$.

First, we make a few remarks on the technical details of the proof. Throughout the proof, we condition on the event $\widetilde{Q}_n = \{\widehat{\boldsymbol{\beta}}_n \in N_n(\delta_n)\}$, where $N_n(\delta_n) = \{\boldsymbol{\beta} \in \mathbb{R}^d : \|(n^{-1}\mathbf{B}_n)^{1/2}(\boldsymbol{\beta} - \boldsymbol{\beta}_{n,0})\|_2 \leq (n/d)^{-1/2}\delta_n\}$, $\mathbf{B}_n = \mathbf{X}^T\text{cov}(\mathbf{Y})\mathbf{X}$, $\delta_n = O(L_n\sqrt{\log p})$ and $\widehat{\boldsymbol{\beta}}_n$ is the unrestricted MLE. Note that the eigenvalues of $n^{-1}\mathbf{A}_n(\boldsymbol{\beta})$ and $n^{-1}\mathbf{B}_n$ are bounded away from 0 and $\infty$ by Conditions 1 and 3. This follows from the fact that eigenvalues of $\mathbf{M}^T\mathbf{N}\mathbf{M}$ lie between $\lambda_{\min}(\mathbf{N})\lambda_{\min}(\mathbf{M}^T\mathbf{M})$ and $\lambda_{\max}(\mathbf{N})\lambda_{\max}(\mathbf{M}^T\mathbf{M})$ for any matrix $\mathbf{M}$ and positive semidefinite symmetric matrix $\mathbf{N}$. Therefore, from Lemma 1 we have that $P(\widetilde{Q}_n) \to 1$ as $n \to \infty$.

To establish this theorem we require a possibly dimension dependent bound on the quantity $\|n^{-1/2}\mathbf{X}\widehat{\boldsymbol{\beta}}_n\|_2$. This can be achieved by putting some restriction on the parameter space. Let $M_n(\alpha) = \{\boldsymbol{\beta} \in \mathbb{R}^d : \|\mathbf{X}\boldsymbol{\beta}\|_\infty \leq \alpha \log n\}$ be a neighborhood, where $\alpha$ is some positive constant. One way of bounding the quantity $\|n^{-1/2}\mathbf{X}\widehat{\boldsymbol{\beta}}_n\|_2$ is to restrict the QMLE $\widehat{\boldsymbol{\beta}}_n$ on the set $M_n(\alpha)$. Here, the constant $\alpha$ can be chosen as large as desired to make $M_n(\alpha)$ large enough, whereas the neighborhood $N_n(\delta_n)$ is asymptotically shrinking. Then, we have $N_n(\delta_n) \subset M_n(\alpha)$ for all sufficiently large $n$, which implies that conditional on $\widetilde{Q}_n$, the restricted MLE coincides with its unrestricted version. Hereafter in this proof $\widehat{\boldsymbol{\beta}}_n$ will be referred to as the restricted MLE, unless specified otherwise.

**Part I: expansion of the term** $\log E_{\mu_{\mathfrak{M}_m}}[U_n(\boldsymbol{\beta})^n]$.

Recall that $U_n(\boldsymbol{\beta}) = \exp\left[n^{-1}\ell_n^*(\mathbf{y},\boldsymbol{\beta})\right]$ and $\ell_n^*(\mathbf{y},\boldsymbol{\beta}) = \ell_n(\mathbf{y},\boldsymbol{\beta}) - \ell_n(\mathbf{y},\widehat{\boldsymbol{\beta}}_n)$. First, we observe that the maximum value of the function $\ell_n^*(\mathbf{y},\boldsymbol{\beta})$ is attained at $\boldsymbol{\beta} = \widehat{\boldsymbol{\beta}}_n$. Moreover, we have $\partial^2 \ell_n^*(\mathbf{y},\boldsymbol{\beta})/\partial \boldsymbol{\beta}^2 = -\mathbf{A}_n(\boldsymbol{\beta})$ from (10) where $\mathbf{A}_n(\boldsymbol{\beta}) = \mathbf{X}^T \boldsymbol{\Sigma}(\mathbf{X}\boldsymbol{\beta})\mathbf{X}$. Then, we consider Taylor's expansion of the log-likelihood function $\ell_n(\mathbf{y},\cdot)$ around $\widehat{\boldsymbol{\beta}}_n$ in a new neighborhood $\widetilde{N}_n(\delta_n) = \{\boldsymbol{\beta} \in \mathbb{R}^d : \|(n^{-1}\mathbf{B}_n)^{1/2}(\boldsymbol{\beta} - \widehat{\boldsymbol{\beta}}_n)\|_2 \leq (n/d)^{-1/2}\delta_n\}$. We get

$$\ell_n^*(\mathbf{y},\boldsymbol{\beta}) = \frac{1}{2}(\boldsymbol{\beta} - \widehat{\boldsymbol{\beta}}_n)^T \left[\partial^2 \ell_n^*(\mathbf{y},\boldsymbol{\beta}_*)/\partial \boldsymbol{\beta}^2\right] (\boldsymbol{\beta} - \widehat{\boldsymbol{\beta}}_n) \tag{30}$$
$$= -\frac{n}{2}\boldsymbol{\delta}^T \mathbf{V}_n(\boldsymbol{\beta}_*)\boldsymbol{\delta},$$

where $\boldsymbol{\beta}_*$ lies on the line segment joining $\boldsymbol{\beta}$ and $\widehat{\boldsymbol{\beta}}_n$, $\boldsymbol{\delta} = n^{-1/2}\mathbf{B}_n^{1/2}(\boldsymbol{\beta} - \widehat{\boldsymbol{\beta}}_n)$, and $\mathbf{V}_n(\boldsymbol{\beta}) = \mathbf{B}_n^{-1/2}\mathbf{A}_n(\boldsymbol{\beta})\mathbf{B}_n^{-1/2}$. Since $\widehat{\boldsymbol{\beta}}_n \in \widetilde{N}_n(\delta_n)$, by the convexity of the neighborhood $\widetilde{N}_n(\delta_n)$ we have $\boldsymbol{\beta}_* \in \widetilde{N}_n(\delta_n)$. We also note that conditional on the event $\widetilde{Q}_n$, it holds that $\widetilde{N}_n(\delta_n) \subset N_n(2\delta_n)$.

Now, we will bound $U_n(\boldsymbol{\beta})^n$ over the region $\widetilde{N}_n(\delta_n)$ using Taylor's expansion in (30). By Condition 5, we get

$$q_1(\boldsymbol{\beta})1_{\widetilde{N}_n(\delta_n)}(\boldsymbol{\beta}) \leq -n^{-1}\ell_n^*(\mathbf{y},\boldsymbol{\beta})1_{\widetilde{N}_n(\delta_n)}(\boldsymbol{\beta}) \leq q_2(\boldsymbol{\beta})1_{\widetilde{N}_n(\delta_n)}(\boldsymbol{\beta}), \tag{31}$$

where $q_1(\boldsymbol{\beta}) = \frac{1}{2}\boldsymbol{\delta}^T[\mathbf{V}_n - \rho_n(\delta_n)\mathbf{I}_d]\boldsymbol{\delta}$ and $q_2(\boldsymbol{\beta}) = \frac{1}{2}\boldsymbol{\delta}^T[\mathbf{V}_n + \rho_n(\delta_n)\mathbf{I}_d]\boldsymbol{\delta}$. Then, we consider the linear transformation $h(\boldsymbol{\beta}) = (n^{-1}\mathbf{B}_n)^{1/2}\boldsymbol{\beta}$. For sufficiently large $n$, we obtain

$$E_{\mu_{\mathfrak{M}}}[e^{-nq_2(\boldsymbol{\beta})}1_{\widetilde{N}_n(\delta_n)}(\boldsymbol{\beta})] \leq E_{\mu_{\mathfrak{M}}}[U_n(\boldsymbol{\beta})^n 1_{\widetilde{N}_n(\delta_n)}(\boldsymbol{\beta})] \leq E_{\mu_{\mathfrak{M}}}[e^{-nq_1(\boldsymbol{\beta})}1_{\widetilde{N}_n(\delta_n)}(\boldsymbol{\beta})], \tag{32}$$

where $\mu_{\mathfrak{M}}$ denotes the prior distribution on $h(\boldsymbol{\beta}) \in \mathbb{R}^d$ for the model $\mathfrak{M}$.

The final expansion of $\log E_{\mu_{\mathfrak{M}}}[U_n(\boldsymbol{\beta})^n]$ results from combination of Lemmas 7–10. The expressions $E_{\mu_{\mathfrak{M}}}[U_n(\boldsymbol{\beta})^n 1_{\widetilde{N}_n^c(\delta_n)}]$ and $\int_{\boldsymbol{\delta} \in \mathbb{R}^d} e^{-nq_j} 1_{\widetilde{N}_n^c(\delta_n)} d\mu_0$ for $j = 1, 2$ in Lemmas 8 and 10 converge to zero faster than any polynomial rate in $n$ since $\kappa_n = \lambda_{\min}(\mathbf{V}_n)/2$ is bounded away from 0. Moreover, Lemmas 7 and 9 yield

$$\log E_{\mu_{\mathfrak{M}}}[U_n(\boldsymbol{\beta})^n] = \log\left\{\left(\frac{2\pi}{n}\right)^{d/2} |\mathbf{V}_n \pm \rho_n(\delta_n)\mathbf{I}_d|^{-1/2}\right\} + \log c_n,$$

where $c_n \in [c_3, c_3^{-1}]$. Finally, we observe that

$$|\mathbf{V}_n \pm \rho_n(\delta_n)\mathbf{I}_d|^{-1/2} = |\mathbf{V}_n|^{-1/2}|\mathbf{I}_d \pm \rho_n(\delta_n)\mathbf{V}_n^{-1}|^{-1/2} = |\mathbf{V}_n|^{-1/2}\{1 + O[\rho_n(\delta_n)\text{tr}(\mathbf{V}_n^{-1})]\}^{-1/2}$$
$$= |\mathbf{V}_n|^{-1/2}\{1 + O[\rho_n(\delta_n)d\lambda_{\min}^{-1}(\mathbf{V}_n)]\}^{-1/2} = |\mathbf{V}_n|^{-1/2}[1 + o(1)],$$

where we use Condition 5. So, we obtain

$$\log E_{\mu_{\mathfrak{M}}}[U_n(\boldsymbol{\beta})^n] = \log\left\{\left(\frac{2\pi}{n}\right)^{d/2} |\mathbf{V}_n|^{-1/2}[1 + o(1)]\right\} + \log c_n$$
$$= -\frac{\log n}{2}d + \frac{1}{2}\log|\mathbf{A}_n^{-1}\mathbf{B}_n| + \frac{\log(2\pi)}{2}d + \log c_n + o(1). \tag{33}$$

This completes the expansion of $\log E_{\mu_{\mathfrak{M}}}[U_n(\boldsymbol{\beta})^n]$.

**Part II: expansion of the prior term** $\log \alpha_{\mathfrak{M}_m}$.

Now, we consider the prior term $\log \alpha_{\mathfrak{M}_m}$ which depends on $\widehat{\boldsymbol{\beta}}_n$ through $D_m$. Simple calculation shows that

$$\log \alpha_{\mathfrak{M}_m} = -D_m + \log C - d \log p. \tag{34}$$

We aim to provide a decomposition of $D_m$ in terms of $\mathbf{H}_n$. Observe that $-D_m = E\eta_n(\widehat{\boldsymbol{\beta}}_n) - \eta_n(\boldsymbol{\beta}_{n,0})$ where $\eta_n(\boldsymbol{\beta}) = E\ell_n(\widetilde{\mathbf{y}}, \boldsymbol{\beta})$, and $\widetilde{\mathbf{y}}$ is an independent copy of $\mathbf{y}$. We expand $E\eta_n(\widehat{\boldsymbol{\beta}}_n)$ around $\eta_n(\boldsymbol{\beta}_{n,0})$. In the GLM setup, we observe that $\ell_n(\widetilde{\mathbf{y}}, \boldsymbol{\beta}) = \widetilde{\mathbf{y}}^T\mathbf{X}\boldsymbol{\beta} - \mathbf{1}^T\mathbf{b}(\mathbf{X}\boldsymbol{\beta})$ and $\eta_n(\boldsymbol{\beta}) = (E\widetilde{\mathbf{y}}^T)\mathbf{X}\boldsymbol{\beta} - \mathbf{1}^T\mathbf{b}(\mathbf{X}\boldsymbol{\beta})$. Then, we split $E\eta_n(\widehat{\boldsymbol{\beta}}_n)$ in the region $\widetilde{Q}_n$ and its complement, that is,

$$\begin{aligned} E\eta_n(\widehat{\boldsymbol{\beta}}_n) &= E\{\eta_n(\widehat{\boldsymbol{\beta}}_n)1_{\widetilde{Q}_n}\} + E\{\eta_n(\widehat{\boldsymbol{\beta}}_n)1_{\widetilde{Q}_n^c}\} \\ &= E\{\eta_n(\widehat{\boldsymbol{\beta}}_n)1_{\widetilde{Q}_n}\} + E\{[(E\widetilde{\mathbf{y}})^T(\mathbf{X}\widehat{\boldsymbol{\beta}}_n) - \mathbf{1}^T\mathbf{b}(\mathbf{X}\widehat{\boldsymbol{\beta}}_n)]1_{\widetilde{Q}_n^c}\}. \end{aligned} \tag{35}$$

First, we aim to show that the second term on the right hand side of (35) is $o(1)$. Performing componentwise Taylor's expansion of $\mathbf{b}(\cdot)$ around $\mathbf{0}$ and evaluating at $\mathbf{X}\widehat{\boldsymbol{\beta}}_n$, we obtain $\mathbf{b}(\mathbf{X}\widehat{\boldsymbol{\beta}}_n) = \mathbf{b}(\mathbf{0}) + b'(0)\mathbf{X}\widehat{\boldsymbol{\beta}}_n + \mathbf{r}$, where $\mathbf{r} = (r_1, \cdots, r_n)^T$ with $r_i = 2^{-1}b''((\mathbf{X}\boldsymbol{\beta}_i^*)_i)(\mathbf{X}\widehat{\boldsymbol{\beta}}_n)_i^2$ and $\boldsymbol{\beta}_1^*, \cdots, \boldsymbol{\beta}_n^*$ lying on the line segment joining $\widehat{\boldsymbol{\beta}}_n$ and $\mathbf{0}$. Thus, we get

$$E\{|(E\widetilde{\mathbf{y}})^T\mathbf{X}\widehat{\boldsymbol{\beta}}_n - \mathbf{1}^T\mathbf{b}(\mathbf{X}\widehat{\boldsymbol{\beta}}_n)|1_{\widetilde{Q}_n^c}\} \leq O\{n\log n + n + n(\log n)^2\}P(\widetilde{Q}_n^c) = o(1) \tag{36}$$

for sufficiently large $n$. The last inequality follows from the fact that $P(\widetilde{Q}_n^c)$ converges to zero faster than any polynomial rate. To verify the orders, we recall that $\widehat{\boldsymbol{\beta}}_n$ is the constrained MLE and $b''(\cdot)$ is bounded away from 0 and $\infty$. Thus, we obtain following bounds for the four terms $|(E\widetilde{\mathbf{y}})^T\mathbf{X}\widehat{\boldsymbol{\beta}}_n| = O(n\log n)$, $|\mathbf{1}^T\mathbf{b}(\mathbf{0})| = O(n)$, $|b'(0)\mathbf{1}^T\mathbf{X}\widehat{\boldsymbol{\beta}}_n| = O(n\log n)$, and $|\mathbf{1}^T\mathbf{r}| = O(n(\log n)^2)$.

Now, we consider the first term on the right hand side of (35). We begin by expanding $\eta_n(\boldsymbol{\beta})$ around $\boldsymbol{\beta}_{n,0}$ conditioned on the event $\widetilde{Q}_n$. By the definition of $\boldsymbol{\beta}_{n,0}$, $\eta_n(\boldsymbol{\beta})$ attains its maximum at $\boldsymbol{\beta}_{n,0}$. By evaluating Taylor's expansion of $\eta_n(\cdot)$ around $\boldsymbol{\beta}_{n,0}$ at $\widehat{\boldsymbol{\beta}}_n$, we derive

$$\begin{aligned} \eta_n(\widehat{\boldsymbol{\beta}}_n) &= \eta_n(\boldsymbol{\beta}_{n,0}) - \frac{1}{2}(\widehat{\boldsymbol{\beta}}_n - \boldsymbol{\beta}_{n,0})^T\mathbf{A}_n(\boldsymbol{\beta}^*)(\widehat{\boldsymbol{\beta}}_n - \boldsymbol{\beta}_{n,0}) \\ &= \eta_n(\boldsymbol{\beta}_{n,0}) - \frac{1}{2}(\widehat{\boldsymbol{\beta}}_n - \boldsymbol{\beta}_{n,0})^T\mathbf{A}_n(\widehat{\boldsymbol{\beta}}_n - \boldsymbol{\beta}_{n,0}) - \frac{s_n}{2}, \end{aligned}$$

where $\mathbf{A}_n(\cdot) = -\partial^2\ell_n(\mathbf{y}, \cdot)/\partial\boldsymbol{\beta}^2$, $\mathbf{A}_n = \mathbf{A}_n(\boldsymbol{\beta}_{n,0})$, and $\boldsymbol{\beta}^*$ is on the line segment joining $\boldsymbol{\beta}_{n,0}$ and $\widehat{\boldsymbol{\beta}}_n$. The second equality is obtained by taking $s_n = (\widehat{\boldsymbol{\beta}}_n - \boldsymbol{\beta}_{n,0})^T[\mathbf{A}_n(\boldsymbol{\beta}^*) - \mathbf{A}_n](\widehat{\boldsymbol{\beta}}_n - \boldsymbol{\beta}_{n,0})$. Furthermore, setting $\mathbf{C}_n = \mathbf{B}_n^{-1/2}\mathbf{A}_n$ and $\mathbf{v}_n = \mathbf{C}_n(\widehat{\boldsymbol{\beta}}_n - \boldsymbol{\beta}_{n,0})$ simplifies the above expression to

$$\eta_n(\widehat{\boldsymbol{\beta}}_n) = \eta_n(\boldsymbol{\beta}_{n,0}) - \frac{1}{2}\mathbf{v}_n^T[(\mathbf{C}_n^{-1})^T\mathbf{A}_n\mathbf{C}_n^{-1}]\mathbf{v}_n - \frac{s_n}{2}. \tag{37}$$

In (37), we first handle the term $s_n$. Note that on the event $\widetilde{Q}_n$, by the convexity of the neighborhood $N_n(\delta_n)$ we have $\boldsymbol{\beta}^* \in N_n(\delta_n)$. Then, Condition 5 implies

$$\left|s_n 1_{\widetilde{Q}_n}\right| = \left|(\widehat{\boldsymbol{\beta}}_n - \boldsymbol{\beta}_{n,0})^T (\mathbf{A}_n(\boldsymbol{\beta}^*) - \mathbf{A}_n)(\widehat{\boldsymbol{\beta}}_n - \boldsymbol{\beta}_{n,0})\right| 1_{\widetilde{Q}_n} \qquad (38)$$
$$= \left|[\mathbf{B}_n^{1/2}(\widehat{\boldsymbol{\beta}}_n - \boldsymbol{\beta}_{n,0})]^T [\mathbf{V}_n(\boldsymbol{\beta}^*) - \mathbf{V}_n][\mathbf{B}_n^{1/2}(\widehat{\boldsymbol{\beta}}_n - \boldsymbol{\beta}_{n,0})]\right| 1_{\widetilde{Q}_n}$$
$$\leq \rho_n(\delta_n) \delta_n^2 d 1_{\widetilde{Q}_n},$$

where $\mathbf{V}_n(\cdot) = \mathbf{B}^{-1/2} \mathbf{A}_n(\cdot) \mathbf{B}_n^{-1/2}$ and $\mathbf{V}_n = \mathbf{V}(\boldsymbol{\beta}_{n,0})$. We then deduce that $E(s_n 1_{\widetilde{Q}_n}) = o(1)$, since $\rho_n(\delta_n) \delta_n^2 d 1_{\widetilde{Q}_n} = o(1)$ by Condition 5. Therefore, (37) becomes

$$E[\eta_n(\widehat{\boldsymbol{\beta}}_n) 1_{\widetilde{Q}_n}] = E[\eta_n(\boldsymbol{\beta}_{n,0}) - \frac{1}{2} \mathbf{v}_n^T [(\mathbf{C}_n^{-1})^T \mathbf{A}_n \mathbf{C}_n^{-1}] \mathbf{v}_n 1_{\widetilde{Q}_n}] + o(1). \qquad (39)$$

We provide a decomposition of $\mathbf{v}_n$ to handle the term $\mathbf{v}_n^T [(\mathbf{C}_n^{-1})^T \mathbf{A}_n \mathbf{C}_n^{-1}] \mathbf{v}_n$ in (39). Define $\boldsymbol{\Psi}(\boldsymbol{\beta}_n) = \mathbf{X}^T [\mathbf{y} - \boldsymbol{\mu}(\mathbf{X} \boldsymbol{\beta}_n)]$. From the score equation we have $\boldsymbol{\Psi}(\widehat{\boldsymbol{\beta}}_n) = 0$. From (8), it holds that $\mathbf{X}^T [E\mathbf{y} - \boldsymbol{\mu}(\mathbf{X}\boldsymbol{\beta}_{n,0})] = 0$. For any $\boldsymbol{\beta}_1, \cdots, \boldsymbol{\beta}_d \in \mathbb{R}^d$, denote by $\widetilde{\mathbf{A}}_n(\boldsymbol{\beta}_1, \cdots, \boldsymbol{\beta}_d)$ a $d \times d$ matrix with $j$th row the corresponding row of $\mathbf{A}_n(\boldsymbol{\beta}_j)$ for each $j = 1, \cdots, d$. Then, we define matrix-valued function $\widetilde{\mathbf{V}}_n(\boldsymbol{\beta}_1, \cdots, \boldsymbol{\beta}_d) = \mathbf{B}_n^{-1/2} \widetilde{\mathbf{A}}_n(\boldsymbol{\beta}_1, \cdots, \boldsymbol{\beta}_d) \mathbf{B}_n^{-1/2}$. Assuming the differentiability of $\boldsymbol{\Psi}(\cdot)$ and applying the mean-value theorem componentwise around $\boldsymbol{\beta}_{n,0}$, we obtain

$$\mathbf{0} = \boldsymbol{\Psi}_n(\widehat{\boldsymbol{\beta}}_n) = \boldsymbol{\Psi}_n(\boldsymbol{\beta}_{n,0}) - \widetilde{\mathbf{A}}_n(\boldsymbol{\beta}_1, \cdots, \boldsymbol{\beta}_d)(\widehat{\boldsymbol{\beta}}_n - \boldsymbol{\beta}_{n,0})$$
$$= \mathbf{X}^T (\mathbf{y} - E\mathbf{y}) - \widetilde{\mathbf{A}}_n(\boldsymbol{\beta}_1, \cdots, \boldsymbol{\beta}_d)(\widehat{\boldsymbol{\beta}}_n - \boldsymbol{\beta}_{n,0}),$$

where each of $\boldsymbol{\beta}_1, \cdots, \boldsymbol{\beta}_d$ lies on the line segment joining $\widehat{\boldsymbol{\beta}}_n$ and $\boldsymbol{\beta}_{n,0}$. Therefore, we have the decomposition

$$\mathbf{v}_n = \mathbf{C}_n(\widehat{\boldsymbol{\beta}}_n - \boldsymbol{\beta}_{n,0}) = \mathbf{u}_n + \mathbf{w}_n, \qquad (40)$$

where $\mathbf{u}_n = \mathbf{B}_n^{-1/2} \mathbf{X}^T (\mathbf{y} - E\mathbf{y})$ and $\mathbf{w}_n = -\left[\widetilde{\mathbf{V}}_n(\boldsymbol{\beta}_1, \cdots, \boldsymbol{\beta}_d) - \mathbf{V}_n\right] \left[\mathbf{B}_n^{1/2}(\widehat{\boldsymbol{\beta}}_n - \boldsymbol{\beta}_{n,0})\right]$.

We handle the quadratic term $\mathbf{v}_n^T [(\mathbf{C}_n^{-1})^T \mathbf{A}_n \mathbf{C}_n^{-1}] \mathbf{v}_n$ in (39) by using the decomposition of $\mathbf{v}_n$. For simplicity of notation, denote by $\mathbf{R}_n = (\mathbf{C}_n^{-1})^T \mathbf{A}_n \mathbf{C}_n^{-1}$. Recall that $\mathbf{C}_n = \mathbf{B}_n^{-1/2} \mathbf{A}_n$. With some calculations we obtain

$$E(\mathbf{u}_n^T \mathbf{R}_n \mathbf{u}_n) = E\{(\mathbf{y} - E\mathbf{y})^T \mathbf{X} \mathbf{A}_n^{-1} \mathbf{X}^T (\mathbf{y} - E\mathbf{y})\}$$
$$= E\{\text{tr}(\mathbf{A}_n^{-1} \mathbf{X}^T (\mathbf{y} - E\mathbf{y})(\mathbf{y} - E\mathbf{y})^T \mathbf{X})\} = \text{tr}(\mathbf{A}_n^{-1} \mathbf{B}_n).$$

Note that $E(\mathbf{u}_n^T \mathbf{R}_n \mathbf{u}_n 1_{\widetilde{Q}_n}) = E(\mathbf{u}_n^T \mathbf{R}_n \mathbf{u}_n) - E(\mathbf{u}_n^T \mathbf{R}_n \mathbf{u}_n 1_{\widetilde{Q}_n^c})$. From Lemma 1, we have $P(\widetilde{Q}_n^c) \to 0$ as $n \to \infty$. We set $\mu_n = \text{tr}(\mathbf{A}_n^{-1} \mathbf{B}_n) \vee 1$, hereby $\mu_n$ is bounded away from zero. We apply Vitali's convergence theorem to show that $E(\mathbf{u}_n^T \mathbf{R}_n \mathbf{u}_n 1_{\widetilde{Q}_n^c}) = o(\mu_n)$. To establish uniform integrability, we use Lemma 6 which states that $\sup_n E|(\mathbf{u}_n^T \mathbf{R}_n \mathbf{u}_n)/\mu_n|^{1+\gamma} < \infty$ for some constant $\gamma > 0$. This leads to $E(\mathbf{u}_n^T \mathbf{R}_n \mathbf{u}_n 1_{\widetilde{Q}_n^c}) = o(\mu_n)$. Hence we have

$$\frac{1}{2} E(\mathbf{u}_n^T \mathbf{R}_n \mathbf{u}_n 1_{\widetilde{Q}_n}) = \frac{1}{2} \text{tr}(\mathbf{A}_n^{-1} \mathbf{B}_n) + o(\mu_n). \qquad (41)$$

Now, it remains to show that

$$E[(\mathbf{w}_n^T\mathbf{R}_n\mathbf{w}_n + 2\mathbf{w}_n^T\mathbf{R}_n\mathbf{u}_n)1_{\widetilde{Q}_n}] = o(\mu_n). \tag{42}$$

Using the definition of $\mathbf{R}_n$ and $\mathbf{w}_n$, we can bound $\mathbf{w}_n^T\mathbf{R}_n\mathbf{w}_n$:

$$\mathbf{w}_n^T\mathbf{R}_n\mathbf{w}_n = \|\mathbf{R}_n^{1/2}\mathbf{w}_n\|_2^2 \le \|\widetilde{\mathbf{V}}_n(\boldsymbol{\beta}_1,\cdots,\boldsymbol{\beta}_d) - \mathbf{V}_n\|_2^2 \delta_n^2 d\,\mathrm{tr}(\mathbf{A}_n^{-1}\mathbf{B}_n).$$

So, on the event $\widetilde{Q}_n$, it holds that $E(\mathbf{w}_n^T\mathbf{R}_n\mathbf{w}_n 1_{\widetilde{Q}_n}) = o(\mu_n)$ by Condition 5. For the cross term $\mathbf{w}_n^T\mathbf{R}_n\mathbf{u}_n$, applying the Cauchy–Schwarz inequality yields

$$|E(\mathbf{w}_n^T\mathbf{R}_n\mathbf{u}_n 1_{\widetilde{Q}_n})| \le E(\|\mathbf{R}_n^{1/2}\mathbf{w}_n\|_2^2 1_{\widetilde{Q}_n})^{1/2} E(\|\mathbf{u}_n^T\mathbf{R}_n^{1/2}\|_2^2)^{1/2}$$
$$\le E[\|\widetilde{\mathbf{V}}_n(\boldsymbol{\beta}_1,\cdots,\boldsymbol{\beta}_d) - \mathbf{V}_n\|_2 1_{\widetilde{Q}_n} \delta_n d^{1/2}\,\mathrm{tr}(\mathbf{A}_n^{-1}\mathbf{B}_n)].$$

Thus, we obtain that $E(\mathbf{w}_n^T\mathbf{R}_n\mathbf{u}_n 1_{\widetilde{Q}_n}) = o(\mu_n)$. Note that $E\{|\eta_n(\boldsymbol{\beta}_{n,0})|1_{\widetilde{Q}_n^c}\}$ is of order $o(1)$ by similar calculations as in (36). Then, combining (35), (39), (41) and (42) yields

$$E\{\eta_n(\widehat{\boldsymbol{\beta}}_n)\} = \eta_n(\boldsymbol{\beta}_{n,0}) - \frac{1}{2}\mathrm{tr}(\mathbf{A}_n^{-1}\mathbf{B}_n) + o(\mu_n). \tag{43}$$

Combining (34) and (43) yields the expansion

$$\log \alpha_{\mathfrak{M}} = -\frac{1}{2}\mathrm{tr}(\mathbf{A}_n^{-1}\mathbf{B}_n) + \log C - d\log p + o(\mu_n).$$

Part I and Part II conclude the proof of Theorem 1.

## A.2 Proof of Theorem 2

In the beginning of the proof, we demonstrate that the theorem follows from the consistency of $\widehat{\mathbf{A}}_n$ and $\widehat{\mathbf{B}}_n$. Next, we establish the consistency of $\widehat{\mathbf{A}}_n$ and $\widehat{\mathbf{B}}_n$. The consistency of $\widehat{\mathbf{A}}_n$ follows directly from the Lipschitz assumption; however, the consistency of $\widehat{\mathbf{B}}_n$ is harder to prove. To accomplish this, we break down $\widehat{\mathbf{B}}_n$ and invoke Bernstein-type tail inequalities and concentration theorems to handle challenging pieces.

We first introduce some notation to simplify the presentation of the proof. $\lambda_k(\cdot)$ denotes the eigenvalues arranged in increasing order. Denote the spectral radius of $d \times d$ square matrix $\mathbf{M}$ by $\rho(\mathbf{M}) = \max_{1\le k\le d}\{|\lambda_k(\mathbf{M})|\}$. $\|\cdot\|_2$ denotes the matrix operator norm. $o_P(\cdot)$ denotes the convergence in probability of the matrix operator norm.

We want to show that $\log|\widehat{\mathbf{H}}_n| = \log|\mathbf{H}_n| + o_P(1)$ and $\mathrm{tr}(\widehat{\mathbf{H}}_n) = \mathrm{tr}(\mathbf{H}_n) + o_P(1)$. To establish both equalities, it is enough to show that $\widehat{\mathbf{H}}_n = \mathbf{H}_n + o_P(1/d)$. Indeed, assume that $\widehat{\mathbf{H}}_n = \mathbf{H}_n + o_P(1/d)$ is established. In that case, we observe that

$$|\mathrm{tr}(\widehat{\mathbf{H}}_n) - \mathrm{tr}(\mathbf{H}_n)| = |\mathrm{tr}(\widehat{\mathbf{H}}_n - \mathbf{H}_n)| \le d\rho(\widehat{\mathbf{H}}_n - \mathbf{H}_n) = d\|\widehat{\mathbf{H}}_n - \mathbf{H}_n\|_2 = o_P(1),$$

where the equality of the spectral radius and the operator norm follows from the symmetry of the matrix $\widehat{\mathbf{H}}_n - \mathbf{H}_n$. Moreover, we have

$$|\log|\widehat{\mathbf{H}}_n| - \log|\mathbf{H}_n|| \leq d \max_{1 \leq k \leq d} |\log \lambda_k(\widehat{\mathbf{H}}_n) - \log \lambda_k(\mathbf{H}_n)|$$

$$= d \max_{1 \leq k \leq d} \log\left(\max\left\{\frac{\lambda_k(\widehat{\mathbf{H}}_n)}{\lambda_k(\mathbf{H}_n)}, \frac{\lambda_k(\mathbf{H}_n)}{\lambda_k(\widehat{\mathbf{H}}_n)}\right\}\right)$$

$$\leq d \max_{1 \leq k \leq d} \left(\max\left\{\frac{\lambda_k(\widehat{\mathbf{H}}_n)}{\lambda_k(\mathbf{H}_n)}, \frac{\lambda_k(\mathbf{H}_n)}{\lambda_k(\widehat{\mathbf{H}}_n)}\right\} - 1\right)$$

$$\leq d \max_{1 \leq k \leq d} \frac{|\lambda_k(\widehat{\mathbf{H}}_n) - \lambda_k(\mathbf{H}_n)|}{\min\{\lambda_k(\widehat{\mathbf{H}}_n), \lambda_k(\mathbf{H}_n)\}}. \tag{44}$$

Recall that the smallest and largest eigenvalues of both $n^{-1}\mathbf{B}_n$ and $n^{-1}\mathbf{A}_n$ are bounded away from 0 and $\infty$. (See the note in the beginning of the proof of Theorem 1.) So, we get $\lambda_k(\mathbf{H}_n) = O(1)$ and $\lambda_k^{-1}(\mathbf{H}_n) = O(1)$ uniformly for all $1 \leq k \leq d$. An application of Weyl's theorem shows that $|\lambda_k(\widehat{\mathbf{H}}_n) - \lambda_k(\mathbf{H}_n)| \leq \rho(\widehat{\mathbf{H}}_n - \mathbf{H}_n)$ for each $k$. We have $\rho(\widehat{\mathbf{H}}_n - \mathbf{H}_n) = \|\widehat{\mathbf{H}}_n - \mathbf{H}_n\|_2 = o_P(1/d)$. Hence, the right hand side of (44) is $o_P(1)$.

Now, we proceed to show that $\widehat{\mathbf{H}}_n = \mathbf{H}_n + o_P(1/d)$. It suffices to prove that $n^{-1}\widehat{\mathbf{A}}_n = n^{-1}\mathbf{A}_n + o_P(1/d)$ and $n^{-1}\widehat{\mathbf{B}}_n = n^{-1}\mathbf{B}_n + o_P(1/d)$. To see the sufficiency, note that

$$\widehat{\mathbf{H}}_n - \mathbf{H}_n = (n^{-1}\widehat{\mathbf{A}}_n)^{-1}(n^{-1}d\widehat{\mathbf{B}}_n) - (n^{-1}\mathbf{A}_n)^{-1}(n^{-1}d\mathbf{B}_n)$$

$$= (n^{-1}\widehat{\mathbf{A}}_n)^{-1}(n^{-1}d\widehat{\mathbf{B}}_n) - (n^{-1}\widehat{\mathbf{A}}_n)^{-1}(n^{-1}d\mathbf{B}_n)$$

$$+ (n^{-1}\widehat{\mathbf{A}}_n)^{-1}(n^{-1}d\mathbf{B}_n) - (n^{-1}\mathbf{A}_n)^{-1}(n^{-1}d\mathbf{B}_n).$$

Then, $\widehat{\mathbf{H}}_n = \mathbf{H}_n + o_P(1/d)$ can be obtained by repeated application of the following properties of the operator norm: $\|(\mathbf{I}_d - \mathbf{M})^{-1}\|_2 \leq 1/(1 - \|\mathbf{M}\|_2)$ if $\|\mathbf{M}\|_2 < 1$, $\|\mathbf{MN}\|_2 \leq \|\mathbf{M}\|_2\|\mathbf{N}\|_2$, and $\|\mathbf{M} + \mathbf{N}\|_2 \leq \|\mathbf{M}\|_2 + \|\mathbf{N}\|_2$, where $\mathbf{M}$ and $\mathbf{N}$ are $d \times d$ matrices [22].

**Part 1: prove $n^{-1}\widehat{\mathbf{A}}_n = n^{-1}\mathbf{A}_n + o_P(1/d)$.** From Lemma 1 we have, $\|\widehat{\boldsymbol{\beta}}_n - \boldsymbol{\beta}_{n,0}\|_2 = O_P\{(n/d)^{-1/2}\delta_n\}$, which entails $\widehat{\boldsymbol{\beta}}_n = \boldsymbol{\beta}_{n,0} + O_P\{(n/d)^{-1/2}\delta_n\}$. Then it follows from the Lipschitz assumption for $n^{-1}\mathbf{A}_n(\boldsymbol{\beta})$ in the neighborhood $N_n(\delta_n)$ that $n^{-1}\widehat{\mathbf{A}}_n = n^{-1}\mathbf{A}_n + o_P(1/d)$.

**Part 2: prove $n^{-1}\widehat{\mathbf{B}}_n = n^{-1}\mathbf{B}_n + o_P(1/d)$.** We need to control the term $\mathbf{y} - \boldsymbol{\mu}(\mathbf{X}\widehat{\boldsymbol{\beta}}_n)$. In correctly specified models, $\boldsymbol{\mu}(\mathbf{X}\boldsymbol{\beta}_{n,0})$ and $E\mathbf{y}$ are the same. So, it is enough to introduce the mean $E\mathbf{y}$ which is close to both $\mathbf{y}$ and $\boldsymbol{\mu}(\mathbf{X}\widehat{\boldsymbol{\beta}}_n)$. However, it is harder to control the term $\mathbf{y} - \boldsymbol{\mu}(\mathbf{X}\widehat{\boldsymbol{\beta}}_n)$ in misspecified models since we need to deal with both $\boldsymbol{\mu}(\mathbf{X}\boldsymbol{\beta}_{n,0})$ and $E\mathbf{y}$.

First, we use the fact that $\boldsymbol{\mu}(\mathbf{X}\boldsymbol{\beta}_{n,0})$ and $\boldsymbol{\mu}(\mathbf{X}\widehat{\boldsymbol{\beta}}_n)$ are close. To accomplish this, we add and subtract $\boldsymbol{\mu}(\mathbf{X}\boldsymbol{\beta}_{n,0})$ to get the following decomposition:

$$n^{-1}\widehat{\mathbf{B}}_n = n^{-1}\mathbf{X}^T \text{diag}\left\{\left[\mathbf{y} - \boldsymbol{\mu}(\mathbf{X}\widehat{\boldsymbol{\beta}}_n)\right] \circ \left[\mathbf{y} - \boldsymbol{\mu}(\mathbf{X}\widehat{\boldsymbol{\beta}}_n)\right]\right\}\mathbf{X}$$

$$= \mathbf{G}_1 + \mathbf{G}_2 + \mathbf{G}_3,$$

where

$$\mathbf{G}_1 = n^{-1}\mathbf{X}^T\mathrm{diag}\{[\mathbf{y} - \boldsymbol{\mu}(\mathbf{X}\boldsymbol{\beta}_{n,0})] \circ [\mathbf{y} - \boldsymbol{\mu}(\mathbf{X}\boldsymbol{\beta}_{n,0})]\}\mathbf{X},$$
$$\mathbf{G}_2 = 2n^{-1}\mathbf{X}^T\mathrm{diag}\{[\mathbf{y} - \boldsymbol{\mu}(\mathbf{X}\boldsymbol{\beta}_{n,0})] \circ [\boldsymbol{\mu}(\mathbf{X}\boldsymbol{\beta}_{n,0}) - \boldsymbol{\mu}(\mathbf{X}\widehat{\boldsymbol{\beta}}_n)]\}\mathbf{X},$$
$$\mathbf{G}_3 = n^{-1}\mathbf{X}^T\mathrm{diag}\{[\boldsymbol{\mu}(\mathbf{X}\widehat{\boldsymbol{\beta}}_n) - \boldsymbol{\mu}(\mathbf{X}\boldsymbol{\beta}_{n,0})] \circ [\boldsymbol{\mu}(\mathbf{X}\widehat{\boldsymbol{\beta}}_n) - \boldsymbol{\mu}(\mathbf{X}\boldsymbol{\beta}_{n,0})]\}\mathbf{X}.$$

Next, we introduce $E\mathbf{y}$ to obtain terms $\mathbf{y} - E\mathbf{y}$ and $E\mathbf{y} - \boldsymbol{\mu}(\mathbf{X}\boldsymbol{\beta}_{n,0})$ both of which can be kept small. We split $\mathbf{G}_1$ as $\mathbf{G}_1 = \mathbf{G}_{11} + \mathbf{G}_{12} + \mathbf{G}_{13}$ and $\mathbf{G}_2$ as $\mathbf{G}_2 = \mathbf{G}_{21} + \mathbf{G}_{22}$, where

$$\mathbf{G}_{11} = n^{-1}\mathbf{X}^T\mathrm{diag}\{(\mathbf{y} - E\mathbf{y}) \circ (\mathbf{y} - E\mathbf{y})\}\mathbf{X},$$
$$\mathbf{G}_{12} = 2n^{-1}\mathbf{X}^T\mathrm{diag}\{(\mathbf{y} - E\mathbf{y}) \circ [E\mathbf{y} - \boldsymbol{\mu}(\mathbf{X}\boldsymbol{\beta}_{n,0})]\}\mathbf{X},$$
$$\mathbf{G}_{13} = n^{-1}\mathbf{X}^T\mathrm{diag}\{[E\mathbf{y} - \boldsymbol{\mu}(\mathbf{X}\boldsymbol{\beta}_{n,0})] \circ [E\mathbf{y} - \boldsymbol{\mu}(\mathbf{X}\boldsymbol{\beta}_{n,0})]\}\mathbf{X},$$
$$\mathbf{G}_{21} = 2n^{-1}\mathbf{X}^T\mathrm{diag}\{(\mathbf{y} - E\mathbf{y}) \circ [\boldsymbol{\mu}(\mathbf{X}\boldsymbol{\beta}_{n,0}) - \boldsymbol{\mu}(\mathbf{X}\widehat{\boldsymbol{\beta}}_n)]\}\mathbf{X},$$
$$\mathbf{G}_{22} = 2n^{-1}\mathbf{X}^T\mathrm{diag}\{[E\mathbf{y} - \boldsymbol{\mu}(\mathbf{X}\boldsymbol{\beta}_{n,0})] \circ [\boldsymbol{\mu}(\mathbf{X}\boldsymbol{\beta}_{n,0}) - \boldsymbol{\mu}(\mathbf{X}\widehat{\boldsymbol{\beta}}_n)]\}\mathbf{X}.$$

Now, we will control each of the above terms separately. Before we begin, we observe that for any matrices $\mathbf{M}$ and $\mathbf{N}$, we have

$$P(d\|\mathbf{M} - \mathbf{N}\|_2 \geq t) \leq P(d\|\mathbf{M} - \mathbf{N}\|_F \geq t)$$
$$\leq d^2 \max_{1 \leq j,k \leq d} P(|\mathbf{M}^{jk} - \mathbf{N}^{jk}| \geq t/d^2), \tag{45}$$

where $\|\cdot\|_F$ denotes the matrix Frobenius norm and $\mathbf{M}^{jk}$ denotes the $(j,k)$th entry of $\mathbf{M}$. Therefore, it is enough to bound $P(|\mathbf{M}^{jk} - \mathbf{N}^{jk}| \geq t/d^2)$ by $o(1/d^2)$ to show that $\mathbf{M} = \mathbf{N} + o_p(1/d)$.

**Part 2a) prove $\mathbf{G}_{11} = n^{-1}\mathbf{B}_n + o_P(1/d)$.** We will use Bernstein-type tail inequality. First, note that $E\mathbf{G}_{11} = n^{-1}\mathbf{B}_n$ and $\mathbf{G}_{11}^{jk} = n^{-1}\sum_{i=1}^n\{x_{ij}x_{ik}[y_i - Ey_i]^2\} = \sum_{i=1}^n a_i^{jk}q_i^2$, where $a_i^{jk} = n^{-1}x_{ij}x_{ik}\mathrm{var}(y_i)$ and $q_i = \{\mathrm{var}(y_i)\}^{-1/2}(y_i - Ey_i)$. Let $\mathbf{a}^{jk} = (a_1^{jk}, \cdots, a_n^{jk})^T$. Then we have $\|\mathbf{a}^{jk}\|_2^2 = O(n^{4u_3 - 1})$ since $\|\mathbf{X}\|_\infty = O(n^{u_3})$ from Condition 3. It may be noted that $q_i$'s are 1-sub-exponential random variables from Condition 1 and so $q_i^2$'s are 2-sub-exponential random variables. Furthermore, $\sup_{1 \leq i \leq n} \mathrm{var}(q_i^2) = O(1)$. To see this, we note

$$\mathrm{var}(q_i^2) \leq Eq_i^4 \leq 4^4(4^{-1}[Eq_i^4]^{1/4})^4 \leq 4^4\left(\sup_{m \geq 1}\{m^{-1}(E|q_i|^m)^{1/m}\}\right)^4 = O(1),$$

where we use Lemma 5. Then combining (45) with Lemma 12 for a choice of $\alpha = 2$, we deduce

$$P(d\|\mathbf{G}_{11} - E\mathbf{G}_{11}\|_2 \geq t) \leq d^2 \max_{1 \leq j,k \leq d} P(|\mathbf{G}_{11}^{jk} - E\mathbf{G}_{11}^{jk}| \geq t/d^2)$$
$$\leq Cd^2\exp\{-Ct^{1/2}n^{\frac{1}{4} - u_3}/d\}$$

for some constant $C$. Since $d = O(n^{\kappa_1})$ and $u < 1/4 - u_3$, the right hand side of above equation tends to zero. Thus, we obtain $\mathbf{G}_{11} = E\mathbf{G}_{11} + o_P(1/d) = n^{-1}\mathbf{B}_n + o_P(1/d)$.

**Part 2b) prove $\mathbf{G}_{12} = o_P(1/d)$.** Similar to the previous part, we invoke Bernstein-type tail inequality. Observe that $\mathbf{G}_{12}^{jk} = n^{-1} \sum_{i=1}^{n} 2\{x_{ij}x_{ik}[E\mathbf{y} - \boldsymbol{\mu}(\mathbf{X}\boldsymbol{\beta}_{n,0})]_i[y_i - Ey_i]\} = \sum_{i=1}^{n} \tilde{a}_i^{jk} q_i$, where $\tilde{a}_i^{jk} = 2n^{-1}\text{var}(y_i)^{1/2}x_{ij}x_{ik}[E\mathbf{y} - \boldsymbol{\mu}(\mathbf{X}\boldsymbol{\beta}_{n,0})]_i$ and $q_i = \{\text{var}(y_i)\}^{-1/2}(y_i - Ey_i)$. Then, we get $\|\tilde{\mathbf{a}}^{jk}\|_2^2 = O(n^{4u_3+u_2/2-3/2})$ by Conditions 2 and 3.

By Lemma 11, we have

$$P(d\|\mathbf{G}_{12}\|_2 \geq t) \leq d^2 \max_{1 \leq j,k \leq d} P(|\mathbf{G}_{12}^{jk}| \geq t/d^2)$$
$$\leq Cd^2 \exp\{-Ctn^{\frac{3}{4}-2u_3-\frac{u_2}{4}}/d^2\}$$

for some constant $C$. Since $d = O(n^{\kappa_1})$ and $3/4 - 2u_3 - u_2/4 - 2\kappa_1 > 0$, the right hand side of above equation tends to zero. Hence, we have $\mathbf{G}_{12} = o_P(1/d)$.

**Part 2c) prove $\mathbf{G}_{13} = o(1/d)$.** We derive

$$\|\mathbf{G}_{13}\|_2^2 \leq \|n^{-1}\sum_{i=1}^{n}\{\mathbf{x}_i\mathbf{x}_i^T[Ey_i - [\boldsymbol{\mu}(\mathbf{X}\boldsymbol{\beta}_{n,0})]_i]^2\}\|_F^2$$
$$= \Sigma_{1 \leq j,k \leq d}[\sum_{i=1}^{n} a_i^{jk}[Ey_i - [\boldsymbol{\mu}(\mathbf{X}\boldsymbol{\beta}_{n,0})]_i]^2/\text{var}(y_i)]^2$$
$$\leq \sum_{i=1}^{n}\{[Ey_i - [\boldsymbol{\mu}(\mathbf{X}\boldsymbol{\beta}_{n,0})]_i]^2/\text{var}(y_i)\}^2 \Sigma_{1 \leq j,k \leq d}\|\mathbf{a}^{jk}\|_2^2,$$

where the last step follows from the componentwise Cauchy–Schwarz inequality. From Conditions 2 and 3, we get $\|\mathbf{G}_{13}\|_2^2 = O(n^{u_2}d^2n^{4u_3-1})$. Therefore, $\mathbf{G}_{13} = o(1/d)$ since $d = O(n^{\kappa_1})$ and $u_2 + 4\kappa_1 + 4u_3 - 1 < 0$.

**Part 2d) prove $\mathbf{G}_{21} = o(1/d^2)$.** Bounding $\mathbf{G}_{21}$ is the trickiest part. The use of classical Bernstein-type inequalities are prohibited since the summation includes two random quantities $\mathbf{y}$ and $\widehat{\boldsymbol{\beta}}$. Instead, we will apply concentration inequalities.

We start by truncating the random variable $y$ by conditioning on the set $\Omega_n = \{\|\mathbf{W}\|_\infty \leq C_1 \log n\}$ which is defined in Lemma 2. Since $\widehat{\boldsymbol{\beta}}_n$ belongs to the neighborhood $N_n(\delta_n)$ by Lemma 1, we get

$$|\mathbf{G}_{21}^{jk}| = |2n^{-1}\sum_{i=1}^{n} x_{ij}x_{ik}[y_i - Ey_i][\boldsymbol{\mu}(\mathbf{X}\boldsymbol{\beta}_{n,0}) - \boldsymbol{\mu}(\mathbf{X}\widehat{\boldsymbol{\beta}}_n)]_i|$$
$$\leq \sup_{\boldsymbol{\beta}_n \in N_n(\delta_n)} 2n^{-1}|\sum_{i=1}^{n} x_{ij}x_{ik}[y_i - Ey_i][\boldsymbol{\mu}(\mathbf{X}\boldsymbol{\beta}_{n,0}) - \boldsymbol{\mu}(\mathbf{X}\boldsymbol{\beta}_n)]_i|.$$

Then, we can separate the right hand side by conditioning on $\Omega_n$. So, we have $|\mathbf{G}_{21}^{jk}| \leq \mathbf{G}_{211}^{jk} + \mathbf{G}_{212}^{jk}$ where

$$\mathbf{G}_{211}^{jk} = \sup_{\boldsymbol{\beta}_n \in N_n(\delta_n)} 2n^{-1}|\sum_{i=1}^{n} x_{ij}x_{ik}[y_i - Ey_i][\boldsymbol{\mu}(\mathbf{X}\boldsymbol{\beta}_{n,0}) - \boldsymbol{\mu}(\mathbf{X}\boldsymbol{\beta}_n)]_i 1_{\Omega_n}|,$$

$$\mathbf{G}_{212}^{jk} = \sup_{\boldsymbol{\beta}_n \in N_n(\delta_n)} 2n^{-1}|\sum_{i=1}^{n} x_{ij}x_{ik}[y_i - Ey_i][\boldsymbol{\mu}(\mathbf{X}\boldsymbol{\beta}_{n,0}) - \boldsymbol{\mu}(\mathbf{X}\boldsymbol{\beta}_n)]_i(1 - 1_{\Omega_n})|.$$

First, we bound $E\mathbf{G}_{211}^{jk}$. We take a Rademacher sequence $\{\epsilon_i\}_{i=1}^n$ independent of $\mathbf{y}$. Then, we apply symmetrization and contraction inequalities in [5] as follows.

$$\begin{aligned}
E\mathbf{G}_{211}^{jk} =& E\sup_{\boldsymbol{\beta}_n\in N_n(\delta_n)} 2n^{-1}|\sum_{i=1}^n x_{ij}x_{ik}[y_i-Ey_i][\boldsymbol{\mu}(\mathbf{X}\boldsymbol{\beta}_{n,0})-\boldsymbol{\mu}(\mathbf{X}\boldsymbol{\beta}_n)]_i 1_{\Omega_n}|\\
\leq & 4n^{-1}E\sup_{\boldsymbol{\beta}_n\in N_n(\delta_n)} |\sum_{i=1}^n \epsilon_i x_{ij}x_{ik}y_i[\boldsymbol{\mu}(\mathbf{X}\boldsymbol{\beta}_{n,0})-\boldsymbol{\mu}(\mathbf{X}\boldsymbol{\beta}_n)]_i 1_{\Omega_n}|\\
\leq & 4n^{-1}c_0 E\sup_{\boldsymbol{\beta}_n\in N_n(\delta_n)} |\sum_{i=1}^n \epsilon_i x_{ij}x_{ik}y_i[\mathbf{X}\boldsymbol{\beta}_{n,0}-\mathbf{X}\boldsymbol{\beta}_n]_i 1_{\Omega_n}|\\
\leq & 4n^{-1}c_0 \sup_{\boldsymbol{\beta}_n\in N_n(\delta_n)} \|\boldsymbol{\beta}_{n,0}-\boldsymbol{\beta}_n\|_2 E\|\sum_{i=1}^n \epsilon_i x_{ij}x_{ik}y_i 1_{\Omega_n}\mathbf{x}_i\|,
\end{aligned}$$

where the last step follows from the Cauchy–Schwarz inequality. We observe that $\sup_{\boldsymbol{\beta}_n\in N_n(\delta_n)} \|\boldsymbol{\beta}_{n,0}-\boldsymbol{\beta}_n\|_2 \leq n^{-1/2}d^{1/2}\delta_n$ and $E\|\sum_{i=1}^n \epsilon_i x_{ij}x_{ik}y_i 1_{\Omega_n}\mathbf{x}_i\|_2 \leq (\sum_{i=1}^n x_{ij}^2 x_{ik}^2 E[y_i^2 1_{\Omega_n}]\|\mathbf{x}_i\|_2^2)^{1/2}$. So, we can bound $E\mathbf{G}_{211}^{jk}$ by $4c_0 n^{-3/2}d^{1/2}\delta_n (\sum_{i=1}^n x_{ij}^2 x_{ik}^2 E[y_i^2 1_{\Omega_n}]\|\mathbf{x}_i\|_2^2)^{1/2}$. Using Conditions 2 and 3, we obtain $E\mathbf{G}_{211}^{jk} = O(n^{-1+2u_3}d\delta_n m_n)$. Since $d=O(n^{\kappa_1})$ and $-1+2u_3+3\kappa_1+2u_1+\kappa_2/2 < 0$, we deduce $E\mathbf{G}_{211}^{jk} = o(1/d^2)$.

Furthermore, we need to bound $2|x_{ij}x_{ik}y_i[\boldsymbol{\mu}(\mathbf{X}\boldsymbol{\beta}_{n,0})-\boldsymbol{\mu}(\mathbf{X}\boldsymbol{\beta}_n)]_i 1_{\Omega_n}|$ for any $\boldsymbol{\beta}_n\in N_n(\delta_n)$ in order to use the concentration theorem in [5]. We use Lemma 2 to bound $y_i$:

$$\begin{aligned}
2|x_{ij}x_{ik}y_i[\boldsymbol{\mu}(\mathbf{X}\boldsymbol{\beta}_{n,0})-\boldsymbol{\mu}(\mathbf{X}\boldsymbol{\beta}_n)]_i 1_{\Omega_n}| \leq & 2|x_{ij}||x_{ik}||(y_i-Ey_i+Ey_i)|1_{\Omega_n}|[\boldsymbol{\mu}(\mathbf{X}\boldsymbol{\beta}_{n,0})-\boldsymbol{\mu}(\mathbf{X}\boldsymbol{\beta}_n)]_i|\\
\leq & 2|x_{ij}||x_{ik}|(|Ey_i|+C_1\log(n))|[\boldsymbol{\mu}(\mathbf{X}\boldsymbol{\beta}_{n,0})-\boldsymbol{\mu}(\mathbf{X}\boldsymbol{\beta}_n)]_i|.
\end{aligned}$$

Since $b''(X\boldsymbol{\beta}) \leq c_0^{-1}$ for any $\boldsymbol{\beta}$ joining the line segment $\boldsymbol{\beta}_{n,0}$ and $\boldsymbol{\beta}_n$, we have $|[\boldsymbol{\mu}(\mathbf{X}\boldsymbol{\beta}_{n,0})-\boldsymbol{\mu}(\mathbf{X}\boldsymbol{\beta}_n)]_i| \leq c_0^{-1}\|\mathbf{x}_i\|_2 \|\boldsymbol{\beta}_{n,0}-\boldsymbol{\beta}_n\|_2$ for any $\boldsymbol{\beta}_n \in N_n(\delta_n)$. When we put last two inequalities together with Conditions 2 and 3, we get $2|x_{ij}x_{ik}y_i[\boldsymbol{\mu}(\mathbf{X}\boldsymbol{\beta}_{n,0})-\boldsymbol{\mu}(\mathbf{X}\boldsymbol{\beta}_n)]_i 1_{\Omega_n}| \leq c_{i,\boldsymbol{\beta}_n}$ where $c_{i,\boldsymbol{\beta}_n} = O(n^{2u_3}m_n)\|\mathbf{x}_i\|_2\|\boldsymbol{\beta}_{n,0}-\boldsymbol{\beta}_n\|_2$. Moreover, we have

$$\begin{aligned}
\sup_{\boldsymbol{\beta}_n\in N_n(\delta_n)} n^{-1}\sum_{i=1}^n c_{i,\boldsymbol{\beta}_n}^2 \leq & O(n^{-1+4u_3}m_n^2) \sup_{\boldsymbol{\beta}_n\in N_n(\delta_n)} \|\boldsymbol{\beta}_{n,0}-\boldsymbol{\beta}_n\|_2^2 \sum_{i=1}^n \|\mathbf{x}_i\|_2^2\\
\leq & O(n^{-1+4u_3}m_n^2 d^2 \delta_n^2)
\end{aligned}$$

where we use the fact that $\|\boldsymbol{\beta}_{n,0}-\boldsymbol{\beta}_n\|_2^2 = O(n^{-1}d\delta_n^2)$ for any $\boldsymbol{\beta}_n \in N_n(\delta_n)$. Thus, we can use the concentration inequality in [5] which yields

$$P(\mathbf{G}_{211}^{jk} \geq E\mathbf{G}_{211}^{jk} + t) \leq C\exp\left\{-C\frac{nt^2}{n^{-1+4u_3}m_n^2 d^2 \delta_n^2}\right\}, \tag{46}$$

for some constant $C$.

Now, take any $\tilde{t} > 0$. We know that $E\mathbf{G}_{211}^{jk} < \tilde{t}/(2d^2)$ for large enough $n$. Then by taking $t = \tilde{t}/(2d^2)$ in equation (46), we obtain

$$P(\mathbf{G}_{211}^{jk} \geq \tilde{t}/d^2) \leq C\exp\{-C\frac{\tilde{t}^2}{n^{-2+4u_3}m_n^2 d^6 \delta_n^2}\}.$$

Since $-2 + 4u_3 + 6\kappa_1 + 4u_1 + \kappa_2 < 0$, we have $P(\mathbf{G}_{211}^{jk} \geq \tilde{t}/d^2) = o(1/d^2)$.

Lastly, $\mathbf{G}_{212}^{jk} = 0$ on the event $\Omega_n$ which holds with probability at least $1 - O(n^{-\delta})$ by Lemma 2. Therefore, we obtain $\mathbf{G}_{21} = o(1/d^2)$ by using (45).

**Part 2e) prove $\mathbf{G}_{22} = o(1/d)$.** First, we apply the Cauchy–Schwarz inequality to obtain

$$|\mathbf{G}_{22}^{jk}|^2 = \left(2\sum_{i=1}^n \left[n^{-1}\mathrm{var}(y_i)^{1/2}x_{ij}x_{ik}[\boldsymbol{\mu}(\mathbf{X}\boldsymbol{\beta}_{n,0}) - \boldsymbol{\mu}(\mathbf{X}\widehat{\boldsymbol{\beta}}_n)]_i\right]\left[\frac{[E\mathbf{y} - \boldsymbol{\mu}(\mathbf{X}\boldsymbol{\beta}_{n,0})]_i}{\mathrm{var}(y_i)^{1/2}}\right]\right)^2$$

$$\leq 4\sum_{i=1}^n n^{-2}\mathrm{var}(y_i)x_{ij}^2x_{ik}^2[\boldsymbol{\mu}(\mathbf{X}\boldsymbol{\beta}_{n,0}) - \boldsymbol{\mu}(\mathbf{X}\widehat{\boldsymbol{\beta}}_n)]_i^2 \sum_{i=1}^n \frac{[E\mathbf{y} - \boldsymbol{\mu}(\mathbf{X}\boldsymbol{\beta}_{n,0})]_i^2}{\mathrm{var}(y_i)}$$

Since $\widehat{\boldsymbol{\beta}}_n$ lies in the region $N_n(\delta_n)$ with high probability and $b''(\cdot)$ is bounded, $[\boldsymbol{\mu}(\mathbf{X}\boldsymbol{\beta}_{n,0}) - \boldsymbol{\mu}(\mathbf{X}\widehat{\boldsymbol{\beta}}_n)]_i^2$ can be bounded by $\|\mathbf{x}_i\|_2^2 O(n^{-1}d\delta_n^2)$. Condition 2 and the Cauchy–Schwarz inequality yield $\sum_{i=1}^n [\mathrm{var}(y_i)]^{-1}[E\mathbf{y} - \boldsymbol{\mu}(\mathbf{X}\boldsymbol{\beta}_{n,0})]_i^2 \leq O(n^{1/2+u_2/2})$. We further use Conditions 1 and 3 to obtain $|\mathbf{G}_{22}^{jk}|^2 = O(n^{-3/2+4u_3+u_2/2}d^2\delta_n^2)$. Since $d = O(n^{\kappa_1})$ and $-3/2 + 4u_3 + u_2/2 + 6\kappa_1 + 2u_1 + \kappa_2 < 0$, we get $|\mathbf{G}_{22}^{jk}|^2 = o(1/d^4)$. Thus, we obtain $\mathbf{G}_{22} = o_p(1/d)$.

**Part 2f) prove $\mathbf{G}_3 = o(1/d)$.** We decompose $(i,j)$th entry of $\mathbf{G}_3$ as follows

$$|\mathbf{G}_3^{jk}| = n^{-1}|\sum_{i=1}^n x_{ij}x_{ik}[\boldsymbol{\mu}(\mathbf{X}\boldsymbol{\beta}_{n,0}) - \boldsymbol{\mu}(\mathbf{X}\widehat{\boldsymbol{\beta}}_n)]_i^2|$$

$$\leq n^{-1}\sum_{i=1}^n |x_{ij}||x_{ik}|[\boldsymbol{\mu}(\mathbf{X}\boldsymbol{\beta}_{n,0}) - \boldsymbol{\mu}(\mathbf{X}\widehat{\boldsymbol{\beta}}_n)]_i^2$$

$$= O(n^{-1+2u_3}d^2\delta_n^2),$$

where the last line is similar to Part 2e. So, $|\mathbf{G}_3^{jk}| = o(1/d^2)$ since $-1 + 2u_3 + 4\kappa_1 + 2u_1 + \kappa_2 < 0$. Therefore, we get $\mathbf{G}_3 = o(1/d)$.

We have finished the proof of Part 2. This concludes the proof of Theorem 2.

## A.3 Proof of Theorem 3

Theorem 3 is a direct consequence of Theorem 2, Lemma 1, and assumption (25). To see this, observe that the difference in the sample version $\mathrm{HGBIC}_p$ can be written as the sum of the population version $\mathrm{HGBIC}_p^*$ and the terms consisting of differences of likelihood, $\mathrm{tr}(\mathbf{H}_n)$ and $\log(\det(\mathbf{H}_n))$ between the sample and population versions. That is,

$$\mathrm{HGBIC}_p(\mathfrak{M}_m) - \mathrm{HGBIC}_p(\mathfrak{M}_1) = \mathrm{HGBIC}_p^*(\mathfrak{M}_m) - \mathrm{HGBIC}_p^*(\mathfrak{M}_1)$$
$$- 2[\ell_n(\mathbf{y}, \widehat{\boldsymbol{\beta}}_{n,m}) - \ell_n(\mathbf{y}, \boldsymbol{\beta}_{n,m,0})] + 2[\ell_n(\mathbf{y}, \widehat{\boldsymbol{\beta}}_{n,1}) - \ell_n(\mathbf{y}, \boldsymbol{\beta}_{n,1,0})]$$
$$+ [\mathrm{tr}(\widehat{\mathbf{H}}_{n,m}) - \mathrm{tr}(\mathbf{H}_{n,m})] - [\mathrm{tr}(\widehat{\mathbf{H}}_{n,1}) - \mathrm{tr}(\mathbf{H}_{n,1})]$$
$$- [\log|\widehat{\mathbf{H}}_{n,m}| - \log|\mathbf{H}_{n,m}|] + [\log|\widehat{\mathbf{H}}_{n,1}| - \log|\mathbf{H}_{n,1}|].$$

The equation (25) suggests that the first line is bounded below by $\Delta$ for any $m > 1$. Then we focus on the remaining terms. Let $m = 2, \cdots, M$ be fixed. The consistency of QMLE in Lemma 1 implies that $-2[\ell_n(\mathbf{y},\widehat{\boldsymbol{\beta}}_{n,m}) - \ell_n(\mathbf{y},\boldsymbol{\beta}_{n,m,0})] + 2[\ell_n(\mathbf{y},\widehat{\boldsymbol{\beta}}_{n,1}) - \ell_n(\mathbf{y},\boldsymbol{\beta}_{n,1,0})]$ converges to zero with probability at least $1 - O(n^{-\delta})$ for some constant $\delta > 0$. Moreover, Theorem 2 proves that the last two lines are also of order $o(\Delta)$ with probability at least $1 - O(n^{-\delta})$. Therefore, $\{\text{HGBIC}_p(\mathfrak{M}_m) - \text{HGBIC}_p(\mathfrak{M}_1)\} > \Delta/2$ with probability $1 - O(n^{-\delta})$ for any fixed $m > 1$. Applying the union bound over all $M = o(n^\delta)$ competing models completes the proof of Theorem 3.

# Supplementary Material to "Large-Scale Model Selection with Misspecification"

Emre Demirkaya, Yang Feng, Pallavi Basu and Jinchi Lv

This Supplementary Material contains key lemmas, their proofs, and additional technical details. All the notation is the same as in the main body of the paper.

## B  Technical lemmas

We aim to establish the asymptotic consistency of QMLE uniformly over all models $\mathfrak{M}$ such that $|\mathfrak{M}| \leq K$ where $K = o(n)$. For this purpose, we extend our notation. $\boldsymbol{\beta}_{n,0}(\mathfrak{M})$ denotes the parameter vector for the working model and is defined by the minimizer of the KL-divergence whose support is $\mathfrak{M}$: $\boldsymbol{\beta}_{n,0}(\mathfrak{M}) = \arg\min_{\boldsymbol{\beta} \in \mathcal{B}(\mathfrak{M})} I(g_n; f_n(\cdot; \boldsymbol{\beta}, \tau))$. $\boldsymbol{\beta}_{n,0}(\mathfrak{M})$ is estimated by the QMLE $\widehat{\boldsymbol{\beta}}(\mathfrak{M})$ which is defined by $\widehat{\boldsymbol{\beta}}(\mathfrak{M}) = \arg\max_{\boldsymbol{\beta} \in \mathcal{B}(\mathfrak{M})} \ell_n(\boldsymbol{\beta})$.

### B.1  Lemma 1 and its proof

**Lemma 1** (Uniform consistency of QMLE). *Assume Conditions 1, 2(i), 3(i), and 3(iii) hold. If $L_n \sqrt{K n^{-1} \log p} \to 0$, then*

$$\sup_{|\mathfrak{M}| \leq K, \mathfrak{M} \subset \{1, \cdots, p\}} \frac{1}{\sqrt{|\mathfrak{M}|}} \|\widehat{\boldsymbol{\beta}}(\mathfrak{M}) - \boldsymbol{\beta}_{n,0}(\mathfrak{M})\|_2 = O_p \left[ L_n \sqrt{n^{-1} \log p} \right],$$

*where $L_n = 2m_n + C_1 \log n$. $m_n$ is a diverging sequence which appears in Condition 2 and $C_1$ is the positive constant from Lemma 2.*

*Proof.* First, we construct the auxiliary parameter vector $\widehat{\boldsymbol{\beta}}_u(\mathfrak{M})$ as follows. For any sequence $N_n$, we take $u = (1 + \|\widehat{\boldsymbol{\beta}}(\mathfrak{M}) - \boldsymbol{\beta}_{n,0}(\mathfrak{M})\|_2/N_n)^{-1}$ and define $\widehat{\boldsymbol{\beta}}_u(\mathfrak{M}) = u\widehat{\boldsymbol{\beta}}(\mathfrak{M}) + (1-u)\boldsymbol{\beta}_{n,0}(\mathfrak{M})$. We have $\|\widehat{\boldsymbol{\beta}}_u(\mathfrak{M}) - \boldsymbol{\beta}_{n,0}(\mathfrak{M})\|_2 = u\|\widehat{\boldsymbol{\beta}}(\mathfrak{M}) - \boldsymbol{\beta}_{n,0}(\mathfrak{M})\|_2 \leq N_n$ by the definition of $u$. So, $\widehat{\boldsymbol{\beta}}_u(\mathfrak{M})$ belongs to the neighborhood $\mathcal{B}_{\mathfrak{M}}(N_n) = \{\boldsymbol{\beta} \in \mathbb{R}^d, \text{supp}(\boldsymbol{\beta}) = \mathfrak{M} : \|\boldsymbol{\beta} - \boldsymbol{\beta}_{n,0}(\mathfrak{M})\|_2 \leq N_n\}$. Moreover, we observe that $\|\widehat{\boldsymbol{\beta}}_u(\mathfrak{M}) - \boldsymbol{\beta}_{n,0}(\mathfrak{M})\|_2 \leq N_n/2$ implies $\|\widehat{\boldsymbol{\beta}}(\mathfrak{M}) - \boldsymbol{\beta}_{n,0}(\mathfrak{M})\|_2 \leq N_n$. Thus, it is enough to bound $\|\widehat{\boldsymbol{\beta}}_u(\mathfrak{M}) - \boldsymbol{\beta}_{n,0}(\mathfrak{M})\|_2$ to prove the theorem.

Now, we consider $\|\widehat{\boldsymbol{\beta}}_u(\mathfrak{M}) - \boldsymbol{\beta}_{n,0}(\mathfrak{M})\|_2$. First, the concavity of $\ell_n$ and the definition of $\widehat{\boldsymbol{\beta}}(\mathfrak{M})$ yield

$$\begin{aligned}
\ell_n(\widehat{\boldsymbol{\beta}}_u(\mathfrak{M})) &\geq u\ell_n(\widehat{\boldsymbol{\beta}}(\mathfrak{M})) + (1-u)\ell_n(\boldsymbol{\beta}_{n,0}(\mathfrak{M})) \\
&\geq u\ell_n(\widehat{\boldsymbol{\beta}}_u(\mathfrak{M})) + (1-u)\ell_n(\boldsymbol{\beta}_{n,0}(\mathfrak{M})).
\end{aligned}$$

So, by rearranging terms, we get

$$-\ell_n(\boldsymbol{\beta}_{n,0}(\mathfrak{M})) + \ell_n(\widehat{\boldsymbol{\beta}}_u(\mathfrak{M})) \geq 0. \tag{A.1}$$

Besides, for any $\boldsymbol{\beta} \in \mathcal{B}_{\mathfrak{M}}(N_n)$, we have

$$E[\ell_n(\boldsymbol{\beta}_{n,0}(\mathfrak{M})) - \ell_n(\boldsymbol{\beta})] = I(g_n; f_n(\cdot; \boldsymbol{\beta}, \tau)) - I(g_n; f_n(\cdot; \boldsymbol{\beta}_{n,0}(\mathfrak{M}), \tau)) \geq 0, \quad (A.2)$$

by the optimality of $\boldsymbol{\beta}_{n,0}(\mathfrak{M})$. Combining (A.1) and (A.2) gives

$$\begin{aligned}
0 &\leq E[\ell_n(\boldsymbol{\beta}_{n,0}(\mathfrak{M})) - \ell_n(\widehat{\boldsymbol{\beta}}_u(\mathfrak{M}))] \\
&\leq -\ell_n(\boldsymbol{\beta}_{n,0}(\mathfrak{M})) + \ell_n(\widehat{\boldsymbol{\beta}}_u(\mathfrak{M})) + E[\ell_n(\boldsymbol{\beta}_{n,0}(\mathfrak{M})) - \ell_n(\widehat{\boldsymbol{\beta}}_u(\mathfrak{M}))] \\
&\leq \sup_{\boldsymbol{\beta} \in \mathcal{B}_{\mathfrak{M}}(N_n)} \left| \ell(\boldsymbol{\beta}) - E[\ell_n(\boldsymbol{\beta})] - \{\ell_n(\boldsymbol{\beta}_{n,0}(\mathfrak{M})) - E[\ell_n(\boldsymbol{\beta}_{n,0}(\mathfrak{M}))]\} \right| \\
&= nT_{\mathfrak{M}}(N_n),
\end{aligned} \quad (A.3)$$

since $\widehat{\boldsymbol{\beta}}_u(\mathfrak{M}) \in \mathcal{B}_{\mathfrak{M}}(N_n)$.

On the other hand, for any $\boldsymbol{\beta} \in \mathcal{B}_{\mathfrak{M}}(N_n)$,

$$\begin{aligned}
E[\ell_n(\boldsymbol{\beta}_{n,0}(\mathfrak{M})) - \ell_n(\boldsymbol{\beta})] &= E\mathbf{Y}^T \mathbf{Z}_{\mathfrak{M}}(\boldsymbol{\beta}_{n,0}(\mathfrak{M}) - \boldsymbol{\beta}) - \mathbf{1}^T (\mathbf{b}(\mathbf{Z}_{\mathfrak{M}} \boldsymbol{\beta}_{n,0}(\mathfrak{M})) - \mathbf{b}(\mathbf{Z}_{\mathfrak{M}} \boldsymbol{\beta})) \\
&= \mu(\mathbf{Z}_{\mathfrak{M}} \boldsymbol{\beta}_{n,0}(\mathfrak{M})) \mathbf{Z}_{\mathfrak{M}}(\boldsymbol{\beta}_{n,0}(\mathfrak{M}) - \boldsymbol{\beta}) - \mathbf{1}^T(\mathbf{b}(\mathbf{Z}_{\mathfrak{M}} \boldsymbol{\beta}_{n,0}(\mathfrak{M})) - \mathbf{b}(\mathbf{Z}_{\mathfrak{M}} \boldsymbol{\beta})),
\end{aligned}$$

since $\boldsymbol{\beta}_{n,0}(\mathfrak{M})$ satisfies the score equation: $\mathbf{Z}_{\mathfrak{M}}^T [E\mathbf{Y} - \mu(\mathbf{Z}_{\mathfrak{M}} \boldsymbol{\beta})] = 0$. Furthermore, applying the second order Taylor expansion yields

$$E[\ell_n(\boldsymbol{\beta}_{n,0}(\mathfrak{M})) - \ell_n(\boldsymbol{\beta})] = \frac{1}{2} \left(\boldsymbol{\beta}_{n,0}(\mathfrak{M}) - \boldsymbol{\beta}\right)^T \mathbf{Z}_{\mathfrak{M}}^T \boldsymbol{\Sigma}(\mathbf{Z}_{\mathfrak{M}} \bar{\boldsymbol{\beta}}) \mathbf{Z}_{\mathfrak{M}} \left(\boldsymbol{\beta}_{n,0}(\mathfrak{M}) - \boldsymbol{\beta}\right),$$

where $\bar{\boldsymbol{\beta}}$ lies on the line segment connecting $\boldsymbol{\beta}_{n,0}(\mathfrak{M})$ and $\boldsymbol{\beta}$. Then, we use Condition 3 and the assumption that $c_0 \leq b''(\mathbf{Z}\boldsymbol{\beta}) \leq c_0^{-1}$ for any $\boldsymbol{\beta} \in \mathcal{B}$. So, we get $E[\ell_n(\boldsymbol{\beta}_{n,0}(\mathfrak{M})) - \ell_n(\boldsymbol{\beta})] \geq \frac{1}{2} n c_0 c_2 \|\boldsymbol{\beta}_{n,0}(\mathfrak{M}) - \widehat{\boldsymbol{\beta}}_u(\mathfrak{M})\|_2^2$. Therefore, for any $\boldsymbol{\beta} \in \mathcal{B}_{\mathfrak{M}}(N_n)$,

$$\|\boldsymbol{\beta}_{n,0}(\mathfrak{M}) - \boldsymbol{\beta}\|_2^2 \leq 2(c_0 c_2)^{-1} n^{-1} E[\ell_n(\boldsymbol{\beta}_{n,0}(\mathfrak{M})) - \ell_n(\boldsymbol{\beta})]. \quad (A.4)$$

Finally, we take a slowly diverging sequence $\gamma_n$ such that $\gamma_n L_n \sqrt{K \log(p)/n} \to 0$. Then, we choose $N_n = \gamma_n L_n \sqrt{|\mathfrak{M}| n^{-1} \log p}$. Since $\widehat{\boldsymbol{\beta}}_u(\mathfrak{M}) \in \mathcal{B}_{\mathfrak{M}}(N_n)$, we combine equations (A.3) and (A.4) to obtain

$$\begin{aligned}
\sup_{|\mathfrak{M}| \leq K} \frac{1}{\sqrt{|\mathfrak{M}|}} \|\boldsymbol{\beta}_{n,0}(\mathfrak{M}) - \widehat{\boldsymbol{\beta}}_u(\mathfrak{M})\|_2 &\leq \sup_{|\mathfrak{M}| \leq K} \left(\frac{T_{\mathfrak{M}}(N_n)}{|\mathfrak{M}|}\right)^{1/2} \sqrt{2(c_0 c_2)^{-1} n^{-1}} \\
&= O_p[L_n \sqrt{n^{-1} \log p}],
\end{aligned}$$

where the last step follows from Lemma 4. This completes the proof of Lemma 1.

## B.2 Lemma 2 and its proof

**Lemma 2.** *Assume that $Y_1, \cdots, Y_n$ are independent and satisfy Condition 1. Then, for any constant $\delta > 0$, there exist large enough positive constants $C_1$ and $C_2$ such that*

$$\|\mathbf{W}\|_\infty \leq C_1 \log n, \quad (A.5)$$

with probability at least $1 - O(n^{-\delta})$ and,

$$\|n^{-1/2} E[\mathbf{W}|\Omega_n]\|_2 = O((\log n)n^{-C_2}), \tag{A.6}$$

where $\Omega_n = \{\|\mathbf{W}\|_\infty \leq C_1 \log n\}$.

*Proof.* We take $t = C_1 \log n$ in Condition 1. So we get

$$P(\|W\|_\infty \leq C_1 \log n) \geq 1 - nP(|W_1| > C_1 \log n) \geq 1 - c_1 n^{1-c_1^{-1}C_1}.$$

We choose $C_1$ large enough so that $1 - c_1^{-1} C_1 \leq 0$. Thus, we have $P(\|W\|_\infty \leq C_1 \log n) = 1 - O(n^{-\delta})$ where we pick $\delta = c_1^{-1} C_1 - 1 > 0$. This proves the first part of the lemma.

Now, we proceed the proof of the second part of the lemma. We will bound each term $E[W_i|\Omega_n]$ for $i = 1, \cdots, n$. Since $\{W_i\}$ for $i = 1, \cdots, n$ are independent, the conditional expectation $E[W_i|\Omega_n]$ can be written as follows

$$E[W_i|\Omega_n] = E[W_i \mid |W_i| \leq C_1 \log n] = \frac{E[W_i 1\{|W_i| \leq C_1 \log n\}]}{P(|W_i| \leq C_1 \log n)}.$$

Since $E\mathbf{W} = 0$ by definition, we get $E[W_i 1\{|W_i| \leq C_1 \log n\}] = -E[W_i 1\{|W_i| > C_1 \log n\}]$. Last two equalities result in

$$|E[W_i|\Omega_n]| \leq \frac{E[|W_i| 1\{|W_i| > C_1 \log n\}]}{P(|W_i| \leq C_1 \log n)}.$$

We already showed that the denominator $P(|W_i| \leq C_1 \log n)$ can be bounded below by $1 - O(n^{-\delta})$ uniformly in $i$. Thus, it suffices to bound the numerator $E[|W_i| 1\{|W_i| > C_1 \log n\}]$. Indeed, we have

$$\begin{aligned}
E[|W_i| 1\{|W_i| > C_1 \log n\}] &= \int_0^\infty P(|W_i| 1\{|W_i| > C_1 \log n\} \geq t) dt \\
&= \int_0^{C_1 \log n} P(|W_i| 1\{|W_i| > C_1 \log n\} \geq t) dt \\
&\quad + \int_{C_1 \log n}^\infty P(|W_i| 1\{|W_i| > C_1 \log n\} \geq t) dt \\
&= \int_0^{C_1 \log n} P(|W_i| \geq C_1 \log n) dt + \int_{C_1 \log n}^\infty P(|W_i| \geq t) dt \\
&\leq C_1 \log n P(|W_i| \geq C_1 \log n) + \int_{C_1 \log n}^\infty c_1 \exp(-c_1^{-1} t) dt \\
&\leq C_1 \log n c_1 \exp(-c_1^{-1} C_1 \log n) + c_1^2 \exp(-c_1^{-1} C_1 \log n),
\end{aligned}$$

where we use Condition 1 in the last two steps. This concludes the proof of Lemma 2 by choosing $C_2 = c_1^{-1} C_1$.

## B.3 Lemma 3 and its proof

**Lemma 3.** *Under Condition 2, the function $\rho$ defined by $\rho(\mathbf{x}_i^T \boldsymbol{\beta}, Y_i) = Y_i \mathbf{x}_i^T \boldsymbol{\beta} - b(\mathbf{x}_i^T \boldsymbol{\beta})$ is Lipschitz continuous with the Lipschitz constant $L_n = 2m_n + C_1 \log n$ conditioned on the set $\Omega_n = \{\|\mathbf{W}\|_\infty \leq C_1 \log n\}$ given in Lemma 2.*

*Proof.* We consider the difference $\rho(\mathbf{x}_i^T \boldsymbol{\beta}_1, Y_i) - \rho(\mathbf{x}_i^T \boldsymbol{\beta}_2, Y_i)$ for any $\boldsymbol{\beta}_1$ and $\boldsymbol{\beta}_2$ in $\mathbb{R}^p$. We observe that

$$|\rho(\mathbf{x}_i^T \boldsymbol{\beta}_1, Y_i) - \rho(\mathbf{x}_i^T \boldsymbol{\beta}_2, Y_i)| \leq |Y_i||\mathbf{x}_i^T (\boldsymbol{\beta}_1 - \boldsymbol{\beta}_2)| + |b(\mathbf{x}_i^T \boldsymbol{\beta}_1) - b(\mathbf{x}_i^T \boldsymbol{\beta}_2)|.$$

We can bound $|Y_i|$ on $\Omega_n$ using Condition 2 as $|Y_i| \leq \|\mathbf{Y}\|_\infty \leq \|E\mathbf{Y}\|_\infty + \|\mathbf{W}\|_\infty \leq m_n + C_1 \log(n)$. Then we apply the mean-value theorem to obtain $|b(\mathbf{x}_i^T \boldsymbol{\beta}_1) - b(\mathbf{x}_i^T \boldsymbol{\beta}_2)| \leq |b'(\tilde{\boldsymbol{\beta}})||\mathbf{x}_i^T(\boldsymbol{\beta}_1 - \boldsymbol{\beta}_2)|$ where $\tilde{\boldsymbol{\beta}}$ lies on the line segment connecting $\boldsymbol{\beta}_1$ and $\boldsymbol{\beta}_2$. Thus, we get $|b(\mathbf{x}_i^T \boldsymbol{\beta}_1) - b(\mathbf{x}_i^T \boldsymbol{\beta}_2)| \leq m_n |\mathbf{x}_i^T(\boldsymbol{\beta}_1 - \boldsymbol{\beta}_2)|$ by Condition 2. Hereby, we showed that $|\rho(\mathbf{x}_i^T \boldsymbol{\beta}_1, Y_i) - \rho(\mathbf{x}_i^T \boldsymbol{\beta}_2, Y_i)| \leq (2m_n + C_1 \log n)|\mathbf{x}_i^T \boldsymbol{\beta}_1 - x_i^T \boldsymbol{\beta}_2|$ conditioned on $\Omega_n$. Thus, $\rho(\cdot, Y_i)$ is Lipschitz continuous with the Lipschitz constant $L_n = 2m_n + C_1 \log n$ conditioned on the set $\Omega_n$. This completes the proof of Lemma 3.

### B.4 Lemma 4 and its proof

**Lemma 4.** *Assume that Conditions 1, 2(i), 3(i), and 3(iii) hold. Define the neighborhood $\mathcal{B}_{\mathfrak{M}}(N) = \{\boldsymbol{\beta} \in \mathbb{R}^d, \operatorname{supp}(\boldsymbol{\beta}) = \mathfrak{M} : \|\boldsymbol{\beta} - \boldsymbol{\beta}_{n,0}(\mathfrak{M})\|_2 \leq N\}$ and*

$$T_{\mathfrak{M}}(N) = \sup_{\boldsymbol{\beta} \in \mathcal{B}_{\mathfrak{M}}(N)} n^{-1} \left|\ell_n(\boldsymbol{\beta}) - \ell_n(\boldsymbol{\beta}_{n,0}(\mathfrak{M})) - E[\ell_n(\boldsymbol{\beta}) - \ell_n(\boldsymbol{\beta}_{n,0}(\mathfrak{M}))]\right|.$$

*If $\gamma_n$ is a slowly diverging sequence such that $\gamma_n L_n \sqrt{Kn^{-1} \log p} \to 0$, then*

$$\sup_{|\mathfrak{M}| \leq K} \frac{1}{|\mathfrak{M}|} T_{\mathfrak{M}}\left(\gamma_n L_n \sqrt{|\mathfrak{M}| n^{-1} \log p}\right) = O(L_n^2 n^{-1} \log p)$$

*with probability at least $(1 - e^2 p^{1-8c_2 \gamma_n^2})(1 - O(n^{-\delta}))$, where $L_n = 2m_n + C_1 \log n$.*

*Proof.* To prove the lemma, we condition on the set $\Omega_n = \{\|Y - EY\|_\infty \leq C_1 \log n\}$. We observe that

$$\begin{aligned}
&\left|\ell_n(\boldsymbol{\beta}) - \ell_n(\boldsymbol{\beta}_{n,0}(\mathfrak{M})) - E[\ell_n(\boldsymbol{\beta}) - \ell_n(\boldsymbol{\beta}_{n,0}(\mathfrak{M}))]\right| \\
&\leq \left|\ell_n(\boldsymbol{\beta}) - \ell_n(\boldsymbol{\beta}_{n,0}(\mathfrak{M})) - E[\ell_n(\boldsymbol{\beta}) - \ell_n(\boldsymbol{\beta}_{n,0}(\mathfrak{M}))|\Omega_n]\right| \\
&\quad + |E[\ell_n(\boldsymbol{\beta}) - \ell_n(\boldsymbol{\beta}_{n,0}(\mathfrak{M}))] - E[\ell_n(\boldsymbol{\beta}) - \ell_n(\boldsymbol{\beta}_{n,0}(\mathfrak{M}))|\Omega_n]|,
\end{aligned}$$

by the triangle inequality. Thus, $T_{\mathfrak{M}}(N_n)$ can be bounded by the sum of the following two terms:

$$\tilde{T}_{\mathfrak{M}}(N_n) = \sup_{\boldsymbol{\beta} \in \mathcal{B}_{\mathfrak{M}}(N_n)} n^{-1} \left|\ell_n(\boldsymbol{\beta}) - \ell_n(\boldsymbol{\beta}_{n,0}(\mathfrak{M})) - E[\ell_n(\boldsymbol{\beta}) - \ell_n(\boldsymbol{\beta}_{n,0}(\mathfrak{M}))|\Omega_n]\right|, \text{ and}$$

$$R_{\mathfrak{M}}(N_n) = \sup_{\boldsymbol{\beta} \in \mathcal{B}_{\mathfrak{M}}(N_n)} n^{-1}\{E[\ell_n(\boldsymbol{\beta}) - \ell_n(\boldsymbol{\beta}_{n,0}(\mathfrak{M}))] - E[\ell_n(\boldsymbol{\beta}) - \ell_n(\boldsymbol{\beta}_{n,0}(\mathfrak{M}))|\Omega_n]\}$$

That is,

$$T_{\mathfrak{M}}(N_n) \leq \tilde{T}_{\mathfrak{M}}(N_n) + R_{\mathfrak{M}}(N_n). \tag{A.7}$$

In the rest of the proof, we will show the following bounds

$$R_{\mathfrak{M}}(N_n) = o\left(L_n^2 \frac{\log p}{n}\right), \tag{A.8}$$

and

$$\tilde{T}_{\mathfrak{M}}(N_n) = O_p\left(L_n^2 \frac{\log p}{n}\right). \tag{A.9}$$

First, we consider $R_{\mathfrak{M}}(N_n)$. We split $R_{\mathfrak{M}}(N_n)$ by the Cauchy–Schwarz inequality so that

$$R_{\mathfrak{M}}(N_n) = \sup_{\boldsymbol{\beta} \in \mathcal{B}_{\mathfrak{M}}(N_n)} n^{-1} |(EY - E[Y|\Omega_n])^T \mathbf{X}[\boldsymbol{\beta} - \boldsymbol{\beta}_{n,0}(\mathfrak{M})]|$$

$$\leq \|n^{-1/2}(EY - E[Y|\Omega_n])\|_2 \sup_{\boldsymbol{\beta} \in \mathcal{B}_{\mathfrak{M}}(N_n)} \|n^{-1/2}\mathbf{X}[\boldsymbol{\beta} - \boldsymbol{\beta}_{n,0}(\mathfrak{M})]\|_2.$$

We have

$$\|n^{-1/2}(EY - E[Y|\Omega_n])\|_2 = \|n^{-1/2}(E[W|\Omega_n])\|_2 = O(n^{-C_2}\log n)$$

by Lemma 2. We also have

$$\|n^{-1/2}\mathbf{X}(\boldsymbol{\beta} - \boldsymbol{\beta}_{n,0}(\mathfrak{M}))\|_2 \leq (\lambda_{\max}(n^{-1}\mathbf{X}_{\mathfrak{M}}^T\mathbf{X}_{\mathfrak{M}}))^{1/2}\|\boldsymbol{\beta} - \boldsymbol{\beta}_{n,0}(\mathfrak{M})\|_2 \leq c_2^{-1/2}N_n,$$

for any $\boldsymbol{\beta} \in \mathcal{B}_{\mathfrak{M}}(N_n)$.

Therefore, $R_{\mathfrak{M}}(\boldsymbol{\beta}) = O(N_n n^{-C_2}\log n)$. So, (A.8) follows by taking $C_2$ large enough.

Next, we deal with the term $\tilde{T}_{\mathfrak{M}}(N_n)$ by showing (A.9). We observe that the difference $\ell_n(\boldsymbol{\beta}) - \ell_n(\boldsymbol{\beta}_{n,0}(\mathfrak{M}))$ can be written as

$$\ell_n(\boldsymbol{\beta}) - \ell_n(\boldsymbol{\beta}_{n,0}(\mathfrak{M})) = \sum_{i=1}^n \left\{ Y_i[\mathbf{x}_i^T\boldsymbol{\beta} - \mathbf{x}_i^T\boldsymbol{\beta}_{n,0}(\mathfrak{M})] - [b(\mathbf{x}_i^T\boldsymbol{\beta}) - b(\mathbf{x}_i^T\boldsymbol{\beta}_{n,0}(\mathfrak{M}))] \right\}$$

$$= \sum_{i=1}^n \left[\rho(\mathbf{x}_i^T\boldsymbol{\beta}, Y_i) - \rho(\mathbf{x}_i^T\boldsymbol{\beta}_{n,0}(\mathfrak{M}), Y_i)\right].$$

In Lemma 3, we showed that $\rho(\mathbf{x}_i^T\boldsymbol{\beta}, Y_i) = Y_i\mathbf{x}_i^T\boldsymbol{\beta} - b(\mathbf{x}_i^T\boldsymbol{\beta})$ is Lipschitz continuous with the Lipschitz constant $L_n$ conditioned on the set $\Omega_n$.

Next, we choose a Rademacher sequence $\{\epsilon_i\}_{i=1}^n$. Then, we apply symmetrization and concentration inequalities in [5] as follows:

$$E[\tilde{T}_{\mathfrak{M}}(N_n)|\Omega_n]$$

$$\leq 2E\left[\sup_{\boldsymbol{\beta} \in \mathcal{B}_{\mathfrak{M}}(N_n)} n^{-1}\left|\sum_{i=1}^n \epsilon_i \left[\rho(\mathbf{x}_i^T\boldsymbol{\beta}, Y_i) - \rho(\mathbf{x}_i^T\boldsymbol{\beta}_{n,0}(\mathfrak{M}), Y_i)\right]\right| \Big| \Omega_n\right]$$

$$\leq 4L_n E\left[\sup_{\boldsymbol{\beta} \in \mathcal{B}_{\mathfrak{M}}(N_n)} n^{-1}\left|\sum_{i=1}^n \epsilon_i(\mathbf{x}_i^T\boldsymbol{\beta} - \mathbf{x}_i^T\boldsymbol{\beta}_{n,0}(\mathfrak{M}))\right| \Big| \Omega_n\right].$$

Furthermore, we have

$$E\left[\sup_{\boldsymbol{\beta}\in\mathcal{B}_{\mathfrak{M}}(N_n)} n^{-1}\left|\sum_{i=1}^n \epsilon_i(\mathbf{x}_i^T\boldsymbol{\beta}-\mathbf{x}_i^T\boldsymbol{\beta}_{n,0}(\mathfrak{M}))\right|\Big|\Omega_n\right]$$

$$\leq E\left[n^{-1}\sup_{\boldsymbol{\beta}\in\mathcal{B}_{\mathfrak{M}}(N_n)}\|\boldsymbol{\beta}-\boldsymbol{\beta}_{n,0}(\mathfrak{M})\|_2\sum_{i=1}^n \epsilon_i(\mathbf{x}_i)_{\mathfrak{M}}\|_2\Big|\Omega_n\right]$$

$$\leq E\left[n^{-1}N_n\|\sum_{i=1}^n \epsilon_i(\mathbf{x}_i)_{\mathfrak{M}}\|_2\Big|\Omega_n\right] = n^{-1}N_n E\left[\left(\sum_{j\in\mathfrak{M}}\left(\sum_{i=1}^n \epsilon_i x_{ij}\right)^2\right)^{1/2}\right]$$

$$\leq n^{-1}N_n \left(\sum_{j\in\mathfrak{M}} E\left[\left(\sum_{i=1}^n \epsilon_i x_{ij}\right)^2\right]\right)^{1/2} = N_n n^{-1/2}|\mathfrak{M}|^{1/2},$$

where we use the Cauchy–Schwarz inequality and the assumption $\sum_{i=1}^n x_{ij}^2 = n$. Therefore, we obtain the bound

$$E[\tilde{T}_{\mathfrak{M}}(N_n)|\Omega_n] \leq 4L_n N_n n^{-1/2}|\mathfrak{M}|^{1/2}. \tag{A.10}$$

For any $\boldsymbol{\beta} \in \mathcal{B}_{\mathfrak{M}}(N_n)$, we have

$$n^{-1}\sum_{i=1}^n |\rho(x_i^T\boldsymbol{\beta}_{n,0}(\mathfrak{M}), Y_i) - \rho(x_i^T\boldsymbol{\beta}, Y_i)|^2$$

$$\leq n^{-1}L_n^2\sum_{i=1}^n |x_i^T\boldsymbol{\beta}_{n,0}(\mathfrak{M}) - x_i^T\boldsymbol{\beta}|^2$$

$$= n^{-1}L_n^2(\boldsymbol{\beta}_{n,0}(\mathfrak{M}) - \boldsymbol{\beta})^T\mathbf{X}_{\mathfrak{M}}^T\mathbf{X}_{\mathfrak{M}}(\boldsymbol{\beta}_{n,0}(\mathfrak{M}) - \boldsymbol{\beta})$$

$$\leq L_n^2 c_2^{-1} N_n^2.$$

Then we apply Theorem 14.2 in [5] to obtain

$$P\left(\tilde{T}_{\mathfrak{M}}(N_n) \geq E[\tilde{T}_{\mathfrak{M}}(N_n)|\Omega_n] + t|\Omega_n\right) \leq \exp\left(\frac{-nc_2 t^2}{8L_n^2 N_n^2}\right).$$

Now, we take $t = 4L_n N_n n^{-1/2}|\mathfrak{M}|^{1/2}u$ for some positive $u$ that will be chosen later. So, we get $P(\tilde{T}_{\mathfrak{M}}(N_n) \geq 4L_n N_n n^{-1/2}|\mathfrak{M}|^{1/2}(1+u)|\Omega_n) \leq \exp(-2c_2 u^2|\mathfrak{M}|)$ by using (A.10).

We choose $N_n = L_n n^{-1/2}|\mathfrak{M}|^{1/2}(1+u)$. So, it follows that

$$P\left(\frac{\tilde{T}_{\mathfrak{M}}(N_n)}{|\mathfrak{M}|} \geq 4L_n^2 n^{-1}(1+u)^2|\Omega_n\right) \leq \exp(-8c_2 u^2|\mathfrak{M}|).$$

Thus, we have

$$P\left(\sup_{|\mathfrak{M}|\leq K}\frac{\tilde{T}_{\mathfrak{M}}(N_n)}{|\mathfrak{M}|} \geq 4L_n^2 n^{-1}(1+u)^2|\Omega_n\right) \leq \sum_{|\mathfrak{M}|\leq K} P\left(\frac{\tilde{T}_{\mathfrak{M}}(N_n)}{|\mathfrak{M}|} \geq 4L_n^2 n^{-1}(1+u)^2|\Omega_n\right)$$

$$\leq \sum_{k\leq K}\binom{p}{k}\exp(-8c_2 u^2 k) \leq \sum_{k\leq K}\left(\frac{pe}{k}\right)^k \exp(-8c_2 u^2 k).$$

Now, we choose $u = \gamma_n \sqrt{\log p}$. So, for $n$ large enough, we get

$$\sum_{k \leq K} \left(\frac{pe}{k}\right)^k \exp(-8c_2 u^2 k) = \sum_{k \leq K} \left(\frac{pe}{k}\right)^k p^{-8c_2 \gamma_n^2 k} = \sum_{k \leq K} \frac{(ep^{(1-8c_2\gamma_n^2)})^k}{k^k}$$

$$\leq \sum_{k \leq K} \frac{ep^{(1-8c_2\gamma_n^2)}}{k!} \leq e^2 p^{1-8c_2\gamma_n^2}.$$

So far, the probability of the event $\tilde{T}_{\mathfrak{M}}(N_n) = O(L_n^2 \log p / n)$, which we call $A$, is bounded below conditional on $\Omega_n$. Simple calculation yields $P(A) \geq P(A \cap \Omega_n) = P(\Omega_n) P(A|\Omega_n)$. Thus, $P(A) \geq (1 - e^2 p^{1-8c_2\gamma_n^2})(1 - O(n^{-\delta}))$. So, (A.9) follows.

We have shown (A.8) and (A.9), which control the terms $\tilde{T}_{\mathfrak{M}}(N_n)$ and $R_{\mathfrak{M}}(N_n)$, respectively. Thus, (A.7) concludes the proof of Lemma 4.

## B.5 Lemma 5 and its proof

**Lemma 5.** *Let $q_i$'s be $n$ independent, but not necessarily identically distributed, scaled and centered random variables with uniform sub-exponential decay, that is,*

$$P(|q_i| > t) \leq C \exp(-C^{-1} t)$$

*for some positive constant $C$. Let $\|q_i\|_{\psi_1}$ denote the sub-exponential norm defined by*

$$\|q_i\|_{\psi_1} := \sup_{m \geq 1} \left\{ m^{-1} (E|q_i|^m)^{1/m} \right\}.$$

*Then, we have $\|q_i\|_{\psi_1} \leq e^{1/e} C(C \vee 1)$ for all $i$.*

*Proof.* From the condition on sub-exponential tails, we derive

$$E|q_i|^m = m \int_0^\infty x^{m-1} P(|q_i| \geq x) dx \leq Cm \int_0^\infty x^{m-1} \exp(-C^{-1} x) dx$$

$$= CmC^m \int_0^\infty u^{m-1} \exp(-u) du = CmC^m \Gamma(m) \leq CmC^m m^m,$$

where the last line follows from the definition of the Gamma function. Taking the $m$th root, we have

$$(E|q_i|^m)^{1/m} \leq (Cm)^{1/m} Cm.$$

Rewriting above equation, we obtain

$$m^{-1}(E|q_i|^m)^{1/m} \leq m^{1/m} C^{1/m} C \leq e^{1/e}(C \vee 1)C,$$

for all $m \geq 1$. Since the bound is independent of $m$, it holds that $\|q_i\|_{\psi_1} \leq e^{1/e} C(C \vee 1)$ for all $i$. This completes the proof of Lemma 5.

## B.6 Lemma 6 and its proof

**Lemma 6.** *Under Condition 1, for some constant $\gamma > 0$, we have*

$$\sup_n E|(\mathbf{u}_n^T \mathbf{R}_n \mathbf{u}_n)/\mu_n|^{1+\gamma} < \infty,$$

*where $\mathbf{u}_n = \mathbf{B}_n^{-1/2}\mathbf{X}^T(\mathbf{Y} - E\mathbf{Y})$, $\mathbf{R}_n = \mathbf{B}_n^{1/2}\mathbf{A}_n^{-1}\mathbf{B}_n^{1/2}$, and $\mu_n = \mathrm{tr}(\mathbf{A}_n^{-1}\mathbf{B}_n) \vee 1$.*

*Proof.* From the expression of $\mathbf{u}_n^T\mathbf{R}_n\mathbf{u}_n$, we have

$$\begin{aligned}\mathbf{u}_n^T\mathbf{R}_n\mathbf{u}_n =& (\mathbf{Y}-E\mathbf{Y})^T\mathbf{X}\mathbf{A}_n^{-1}\mathbf{X}^T(\mathbf{Y}-E\mathbf{Y})\\=& [(\mathbf{Y}-E\mathbf{Y})^T\mathrm{cov}(\mathbf{Y})^{-1/2}][\mathrm{cov}(\mathbf{Y})^{1/2}\mathbf{X}\mathbf{A}_n^{-1}\mathbf{X}^T\mathrm{cov}(\mathbf{Y})^{1/2}]\\&\cdot [\mathrm{cov}(\mathbf{Y})^{-1/2}(\mathbf{Y}-E\mathbf{Y})].\end{aligned}$$

Denote $\mathbf{S}_n = \mathrm{cov}(\mathbf{Y})^{1/2}\mathbf{X}\mathbf{A}_n^{-1}\mathbf{X}^T\mathrm{cov}(\mathbf{Y})^{1/2}$ and $\mathbf{q} = \mathrm{cov}(\mathbf{Y})^{-1/2}(\mathbf{Y} - E\mathbf{Y})$. We decompose $\mathbf{u}_n^T\mathbf{R}_n\mathbf{u}_n$ into two terms, the summations of the diagonal entries and the off-diagonal entries, respectively,

$$\mathbf{u}_n^T\mathbf{R}_n\mathbf{u}_n = \mathbf{q}^T\mathbf{S}_n\mathbf{q} = \sum_{i=1}^n s_{ii}q_i^2 + \sum_{1 \le i \ne j \le n} s_{ij}q_iq_j,$$

where $s_{ij}$ and $q_i$ denote the $(i,j)$th entry of $\mathbf{S}_n$ and $i$th entry of $\mathbf{q}$. Then, we have

$$\begin{aligned}E(\mathbf{u}_n^T\mathbf{R}_n\mathbf{u}_n)^2 =& \sum_{i=1}^n s_{ii}^2 E(q_i^4) + \sum_{1 \le i \ne j \le n} s_{ii}s_{jj}E(q_i^2)E(q_j^2) \\&+ 2\sum_{1 \le i \ne j \le n} s_{ij}^2 E(q_i^2)E(q_j^2).\end{aligned}$$

Using Condition 1 and the sub-Gaussian norm bound in Lemma 5, both quantities $E(q_i^4)$ and $E(q_i^2)E(q_j^2)$ can be uniformly bounded by a common constant. Hence

$$E(\mathbf{u}_n^T\mathbf{R}_n\mathbf{u}_n)^2 \le O(1) \cdot \{[\mathrm{tr}(\mathbf{S}_n)]^2 + \mathrm{tr}(\mathbf{S}_n^2)\}.$$

Since $\mathbf{S}_n$ is positive semidefinite it holds that $\mathrm{tr}(\mathbf{S}_n^2) \le [\mathrm{tr}(\mathbf{S}_n)]^2$. Finally noting that $\mathrm{tr}(\mathbf{S}_n) = \mathrm{tr}(\mathbf{A}_n^{-1}\mathbf{B}_n) \le \mu_n$, we see that $\sup_n E|(\mathbf{u}_n^T\mathbf{R}_n\mathbf{u}_n)/\mu_n|^{1+\gamma} < \infty$ for $\gamma = 1$, which concludes the proof of Lemma 6.

## C  Additional technical details

Lemmas 7–10 below are similar to those in [28]. Their proofs can be found in [28] or with minor modifications.

**Lemma 7.** *Under Condition 4, for $j = 1, 2$, we have*

$$c_4 \int_{\boldsymbol{\delta}\in\mathbb{R}^d} e^{-nq_j} 1_{\widetilde{N}_n(\delta_n)} d\mu_0 \le E_{\mu_\mathfrak{M}}\left[e^{-nq_j}1_{\widetilde{N}_n(\delta_n)}\right] \le c_5 \int_{\boldsymbol{\delta}\in\mathbb{R}^d} e^{-nq_j} 1_{\widetilde{N}_n(\delta_n)} d\mu_0. \quad (A.11)$$

**Lemma 8.** *Conditional on the event $\widetilde{Q}_n$, for sufficiently large $n$ we have*

$$E_{\mu_{\mathfrak{M}}}[U_n(\boldsymbol{\beta})^n 1_{\widetilde{N}_n^c(\delta_n)}] \leq \exp\{-[\kappa_n - \rho_n(\delta_n)/2]d\delta_n^2\} \qquad \text{(A.12)}$$
$$\leq \exp[-(\kappa_n/2)d\delta_n^2],$$

*where $\kappa_n = \lambda_{\min}(\mathbf{V}_n)/2$.*

**Lemma 9.** *It holds that*

$$\int_{\boldsymbol{\delta}\in\mathbb{R}^d} e^{-nq_1} d\mu_0 = \left(\frac{2\pi}{n}\right)^{d/2} |\mathbf{V}_n - \rho_n(\delta_n)\mathbf{I}_d|^{-1/2} \qquad \text{(A.13)}$$

*and*

$$\int_{\boldsymbol{\delta}\in\mathbb{R}^d} e^{-nq_2} d\mu_0 = \left(\frac{2\pi}{n}\right)^{d/2} |\mathbf{V}_n + \rho_n(\delta_n)\mathbf{I}_d|^{-1/2}. \qquad \text{(A.14)}$$

**Lemma 10.** *For $j = 1, 2$, it holds that*

$$\int_{\boldsymbol{\delta}\in\mathbb{R}^d} e^{-nq_j} 1_{\widetilde{N}_n^c(\delta_n)} d\mu_0 \leq \left(\frac{2\pi}{n\kappa_n}\right)^{d/2} \exp\left[-(\sqrt{\kappa_n d\delta_n^2} - \sqrt{d})^2/2\right]. \qquad \text{(A.15)}$$

**Lemma 11** ([35]). *For independent sub-exponential random variables $\{y_i\}_{i=1}^n$, we have that the sub-exponential norm of $q_i = \{\mathrm{var}(y_i)\}^{-1/2}(y_i - Ey_i)$ is bounded by some positive constant $C_3$. Moreover, the following Bernstein-type tail probability bound holds*

$$P\left\{|\sum_{i=1}^n a_i q_i| \geq t\right\} \leq 2\exp\left[-C_3 \min\left(\frac{t^2}{C_3^2 \|\mathbf{a}\|_2^2}, \frac{t}{C_3 \|\mathbf{a}\|_\infty}\right)\right]$$

*for $\mathbf{a} \in \mathbb{R}^n$, $t \geq 0$.*

Lemma 11 rephrases Proposition 5.16 of [35] for the case where $\|q_i\|_{\Psi_1} \leq C_3$. Further, for our proof we need to characterize the concentration of the square of a sub-exponential random variable. In this regard, we define a general $\alpha$-sub-exponential random variable $\xi_\alpha$ which satisfies

$$P(|\xi_\alpha| > t^\alpha) \leq H\exp(-t/H)$$

for $H, t > 0$. Note that the usual sub-exponential $q_i$'s are 1-sub-exponential random variables. It may be useful to note that $\alpha = 1/2$ corresponds to sub-Gaussian random variables.

**Lemma 12** ([12]). *For independent $\alpha$-sub-exponential random variables $q_i^2$, the following Bernstein-type tail probability bound holds*

$$P\left\{|\sum_{i=1}^n a_i q_i^2 - E[\sum_{i=1}^n a_i q_i^2]| \geq t\right\} \leq C_4 \exp\left[-C_4 \left(\frac{t}{\sup_i \mathrm{var}^{1/2}(q_i^2)\|\mathbf{a}\|_2}\right)^{\frac{2}{2+\alpha}}\right]$$

*for $\mathbf{a} \in \mathbb{R}^n$, $t \geq \sup_i \mathrm{var}^{1/2}(q_i^2)\|\mathbf{a}\|_2$, and $C_4 > 0$ depending on the choice of $\alpha, H$.*

The proof of Lemma 12 follows from that of Lemma 8.2 in [12].